# Quasinormal mode solvers for resonators with dispersive materials

P. Lalanne[1]\*, W. Yan[1], A. Gras[1,10], C. Sauvan[2], J.-P. Hugonin[2], M. Besbes[2], G. Demésy[3], M. D. Truong[3], B. Gralak[3], F. Zolla[3], A. Nicolet[3], F. Binkowski[4], L. Zschiedrich[5], S. Burger[4-5], J. Zimmerling[6], R. Remis[6], P. Urbach[7], H. T. Liu[8], T. Weiss[9]

[1]LP2N, Institut d'Optique Graduate School, CNRS, Univ. Bordeaux, 33400 Talence, France

[2]Laboratoire Charles Fabry, Institut d'Optique Graduate School, CNRS, Université Paris-Saclay, 91127 Palaiseau Cedex, France

[3]Aix Marseille Univ, CNRS, Centrale Marseille, Institut Fresnel, Marseille, France

[4]Zuse Institute Berlin, 14195 Berlin, Germany,

[5]JCMwave GmbH, 14050 Berlin, Germany

[6]Circuits and Systems, Delft University of Technology,  Mekelweg 2, 2628 CD Delft, The Netherlands

[7]Optics Research Group, Delft University of Technology,  Mekelweg 2, 2628 CD Delft, The Netherlands

[8]Tianjin Key Laboratory of Optoelectronic Sensor and Sensing Network Technology, Institute of Modern Optics, College of Electronic Information and Optical Engineering, Nankai University, Tianjin 300350, China

[9]Physics Institute and Research Center SCoPE, University of Stuttgart, Pfaffenwaldring 57, 70550 Stuttgart, Germany

\* E-mail: philippe.lalanne@institutoptique.fr

## Abstract

Optical resonators are widely used in modern photonics. Their spectral response and temporal dynamics are fundamentally driven by their natural resonances, the so-called quasinormal modes (QNMs), with complex frequencies. For optical resonators made of dispersive materials, the QNM computation requires solving a nonlinear eigenvalue problem. This rises a difficulty that is only scarcely documented in the literature. We review our recent efforts for implementing efficient and accurate QNM-solvers for computing and normalizing the QNMs of micro- and nano-resonators made of highly-dispersive materials. We benchmark several methods for three geometries, a two-dimensional plasmonic crystal, a two-dimensional metal grating, and a three-dimensional nanopatch antenna on a metal substrate, in the perspective to elaborate standards for the computation of resonance modes.

**KEYWORDS**: electromagnetic resonance, quasinormal mode, microcavity, nanoresonator, nanoantenna, plasmonic crystals.

## 1. Introduction

Optical resonances play an essential role in current developments in nanophotonics. By providing an eagle view of the resonant features of nanostructures, they are at the heart of the design of artificial materials, integrated photonic resonators, optical sensors, and nanoparticle traps for instance, and find use in many areas of science and technologies.

Modes in optics are defined as factorized solutions $\left[ \tilde{\mathbf{E}}_m(\mathbf{r}), \tilde{\mathbf{H}}_m(\mathbf{r}) \right] \exp(-i\tilde{\omega}_m t)$ of the time-harmonic source-

free Maxwell's equations [Lal18]

$$\begin{bmatrix} 0 & i\varepsilon^{-1}(\mathbf{r},\widetilde{\omega}_m)\nabla\times \\ -i\mu^{-1}(\mathbf{r},\widetilde{\omega}_m)\nabla\times & 0 \end{bmatrix}\begin{bmatrix} \tilde{\mathbf{E}}_m(\mathbf{r}) \\ \tilde{\mathbf{H}}_m(\mathbf{r}) \end{bmatrix} = \widetilde{\omega}_m\begin{bmatrix} \tilde{\mathbf{E}}_m(\mathbf{r}) \\ \tilde{\mathbf{H}}_m(\mathbf{r}) \end{bmatrix}, \tag{1}$$

where the electromagnetic field satisfies the far-field outgoing wave conditions at infinity (Sommerfeld conditions for instance in uniform media), $\boldsymbol{\varepsilon}(\mathbf{r},\omega)$ and $\boldsymbol{\mu}(\mathbf{r},\omega)$ are the position and frequency-dependent permittivity and permeability tensors of the resonator and its surrounding background. Equation (1) takes the form of an eigenproblem, with $\widetilde{\omega}_m$ and $\left[\tilde{\mathbf{E}}_m(\mathbf{r}),\tilde{\mathbf{H}}_m(\mathbf{r})\right]^T$ being the eigenvalues and eigenvectors, respectively.

There are different types of modes in optics. For optical (lossless) dielectric waveguides, truly-guided modes have a *real* (angular) frequency $\omega$, implying that the mode lifetimes are infinite. These modes that live forever in the waveguide are not considered in this work, since efficient mode solvers already exist for their computation. In the presence of leakage *or* material Joule losses, the eigenvalue problem of Eq. (1) becomes non-Hermitian and its eigenvalues lie in the complex plane (lower half plane, in the case of the $e^{-i\omega t}$ time dependence that is assumed throughout the paper)

$$\widetilde{\omega}_m = \Omega_m - i\Gamma_m/2, \tag{2}$$

where the real and imaginary parts give the resonance frequency and the damping rate, or the mode lifetime $\tau_m = 1/\Gamma_m$. In a spectrum, small (resp. large) values of $\Gamma_m$ therefore correspond to narrow (resp. broad) resonances. The damping has two origins, absorption (Joule loss) or leakage into the claddings for open systems. Realistic resonant systems usually consist of resonators connected to the outside world through one or several radiative channels and thus, even in the absence of absorption, $\Gamma_m \neq 0$. To emphasize their difference with the normal modes of Hermitian systems (without any damping), the modes of non-Hermitian systems with complex eigenfrequencies are called quasinormal modes (QNMs) [Cha96,Lal18]. These modes are also known in the literature as decaying states [Mor71], resonant states [Mor73,Mul10], leaky modes [Sny83], scattering modes…

QNMs are initially loaded by a driving field and then decay exponentially with time due to power leakage and absorption. In general, one expects that the electromagnetic field $\left[\mathbf{E}(\mathbf{r},t),\mathbf{H}(\mathbf{r},t)\right]$ scattered by a resonator driven by an external source can be expanded over the QNMs of the system, $\left[\mathbf{E}(\mathbf{r},t),\mathbf{H}(\mathbf{r},t)\right] = \mathrm{Re}\{\sum_m \beta_m(t)\exp(-i\Omega_m t)\exp(-\Gamma_m t/2)\left[\tilde{\mathbf{E}}_m(\mathbf{r}),\tilde{\mathbf{H}}_m(\mathbf{r})\right]\}$, with $\beta_m$ the modal excitation coefficient. Recovering the resonator response in the modal basis is called the reconstruction problem. Modal methods are important as they help interpreting experimental results and highlighting the physics [Lal18]. In this paper, we do not consider the reconstruction problem but focus on the first essential step: the computation of normalized QNMs. We benchmark different techniques by considering three different examples.

It is convenient to consider step by step the impact of damping on the mode computation. When the damping is only due to absorption (there is no leakage), the system is closed. This case is the simplest one and is studied in the first benchmarked structure, a two-dimensional (2D) plasmonic crystal composed of a periodic array of metallic wires. The complexity of this example comes from the dispersive nature of the resonator material, which makes the eigenproblem of Eq. (1) a nonlinear one. Note that for non-dispersive materials, the eigenproblem is linear and efficient QNM solvers already exist in several commercial software packages.

For the second benchmark, we are again concerned by a 2D geometry, but an open one that is periodic in one direction only: a grating composed of an array of slits etched into a metal membrane suspended in air. QNMs now have to fulfill the outgoing-wave boundary conditions for $|\mathbf{r}| \to \infty$ to ensure that the QNM energy leaks away from the resonator. The open nature of the eigenproblem results in an unusual yet critical feature of QNMs, being that the field distributions $\left[\widetilde{\mathbf{E}}_m(\mathbf{r}), \widetilde{\mathbf{H}}_m(\mathbf{r})\right]$ diverge outside the resonator as $|\mathbf{r}| \to \infty$. Indeed, the temporal response imposes an exponential decay for $\exp[-i\widetilde{\omega}_m(t - r/c)]$ and thus a negative imaginary part for $\widetilde{\omega}_m$. As a consequence, an outgoing wave of the form $\exp[-i\widetilde{\omega}_m(t - r/c)]/r$, as encountered in the far field of the resonator, grows exponentially as $\exp[\Gamma_m r/(2c)]/r$ because of the necessary minus sign of the causal propagation term $(t - r/c)$. The exponential divergence has raised problems and even debates in the past for normalizing the QNMs, but this issue is solved nowadays, even for complicated 3D geometries [Lal18]. Normalization is a key point of QNM computation, since only normalized fields can be used to define the modal excitation coefficients of the reconstruction problem. Hereafter, all the computed QNMs are normalized and the convergence of the methods towards the normalized fields is systematically considered in the benchmarks.

The last benchmark is related to a 3D geometry, a silver nanocube deposited on a gold substrate. With a tiny dielectric gap separating the two metals, the geometry offers deep-subwavelength confinements at visible frequencies, for which we compute the fundamental magnetic-dipole mode. With two different metals, this is the most challenging test case. As a simpler version, we also consider an axisymmetric geometry with a nanocylinder.

The article is divided into four additional sections. Section 2 provides a short theoretical background on the normalization of QNMs and an overview of the different techniques used to compute QNMs in electromagnetism. The purpose of this section is to provide the reader with sufficient background in order to obtain a basic understanding of the numerical approaches used by the different research groups involved in the benchmarks. In Section 3, we summarize the main results obtained for the three benchmarks and compare the numerical results obtained by the different groups. We additionally discuss the discrepancies between the results of the different methods. Details concerning the implementations of the methods are described in Section 4 together with additional data and related literature. We include, however, a description of all specific modifications made in order to improve the performance of the methods for the considered geometry. Wall times and memory requirements are likewise provided in this section. We conclude this paper in Section 5 with a summary of the derived insights.

## 2. Quasinormal mode normalization and mode volume

### 2.1 Reconstruction

An important objective, not comprehensively discussed in this article, of QNM theories consists in reconstructing the field $[\mathbf{E}_S(\mathbf{r}, \omega), \mathbf{H}_S(\mathbf{r}, \omega)] \exp(-i\omega t)$ scattered by a resonator (at least in a compact subspace of $\mathbb{R}^3$) under a driving field at a real frequency $\omega$ with a QNM expansion of the form

$$\begin{bmatrix} \mathbf{E}_S(\mathbf{r}, \omega) \\ \mathbf{H}_S(\mathbf{r}, \omega) \end{bmatrix} = \sum_m \alpha_m(\omega) \begin{bmatrix} \widetilde{\mathbf{E}}_m(\mathbf{r}) \\ \widetilde{\mathbf{H}}_m(\mathbf{r}) \end{bmatrix}, \tag{3}$$

in the frequency domain, or

$$\begin{bmatrix} \mathbf{E}_S(\mathbf{r}, t) \\ \mathbf{H}_S(\mathbf{r}, t) \end{bmatrix} = \mathrm{Re}\left( \sum_m \beta_m(t) \begin{bmatrix} \widetilde{\mathbf{E}}_m(\mathbf{r}) \\ \widetilde{\mathbf{H}}_m(\mathbf{r}) \end{bmatrix} \right), \tag{4}$$

in the time domain, provided that the excitation pulse can be Fourier transformed. In Eqs. (3) and (4), the $\alpha_m$'s and $\beta_m$'s are complex modal excitation coefficients, which measure how much the QNMs are excited. There are different analytical formulas to compute the $\alpha_m$'s and the $\beta_m$'s, see [Lal18] for a review. Note that throughout the manuscript, we will use a tilde to differentiate the QNM fields from other fields, for instance the scattered or driving fields. Consistently, we will also use a tilde to denote the QNM complex frequency $\tilde{\omega}_m$, $m = 1,2 \ldots$, in contrast with the real excitation frequencies that will be denoted by $\omega$. The intrinsic strength of QNM expansions is to provide important clues towards understanding the physics of the resonator response. Since the modes are explicitly considered, their impact on the resonance is readily available and unambiguous, in sharp contrast with classical scattering theories.

## 2.2 Mode normalization

The modal excitation coefficients contained in expansions such as those of Eqs. (3) or (4) are functions of normalized QNMs [Lal18]. Normalization has been a long standing issue because of the spatially exponential divergence of the QNM field. Initial works on QNM normalization [Leu94] focused on simple geometries for which the field is known analytically in the whole space, i.e., 1D Fabry-Perot cavities and spheres. For resonators with complex shapes and materials, analytic solutions are not available and the continuous Maxwell's operator of Eq. (1) has to be approximated and expressed using a numerical discretization scheme, which preserves the outgoing-wave condition at $|\boldsymbol{r}| \to \infty$. For a long time, the norm

$$\iiint_\Omega \left( \varepsilon |\tilde{\mathbf{E}}_m|^2 + \mu |\tilde{\mathbf{H}}_m|^2 \right) d^3\mathbf{r}, \tag{5}$$

which comes from Hermitian (closed) systems and represents the field energy, has been used [Ger03] to normalize QNMs of resonators with a high quality factor $Q$. In practice, the integration domain $\Omega$ has to be truncated to some appropriate volume, which is large enough to include most of the physically-reasonable field energy and small enough not to feel the spatial divergence of the QNM field. This approximate way is only valid in the limit of infinitely large $Q$'s. However, for low-$Q$ resonators such as plasmonic antennas, Eq. (5) becomes incorrect, and the classical energy normalization must be replaced by [Sau13]

$$\iiint_\Omega \left[ \tilde{\mathbf{E}}_m \cdot \frac{\partial(\omega\varepsilon)}{\partial\omega} \tilde{\mathbf{E}}_m - \tilde{\mathbf{H}}_m \cdot \frac{\partial(\omega\mu)}{\partial\omega} \tilde{\mathbf{H}}_m \right] d^3\mathbf{r} = 1. \tag{6}$$

The replacement of modulus-squared terms, e.g. $|\tilde{\mathbf{E}}_m|^2$ in Eq. (5), by unconjugated scalar products, e.g. $\tilde{\mathbf{E}}_m \cdot \tilde{\mathbf{E}}_m$ in Eq. (6), enable convergence, as we replace a positive exponentially diverging quantity by an oscillating exponentially diverging quantity that is alternatively positive and negative in space. However, convergence is not guaranteed and, without any particular precautions being taken, the integral defined by Eq. (6) is also diverging if the integration is performed over the entire space [Mu16b].

Actually, it is possible to show that Eq. (6) defines a normalization by surrounding the resonator with perfectly-matched layers (PMLs) [Sau13]. PMLs are often seen as a numerical trick, but they are mainly a mathematically powerful tool: complex coordinate transforms implemented by changing material parameters. The PML mapping offers a precious advantage. The mapped QNMs do not grow exponentially away from the resonator; instead they are even exponentially damped inside the PMLs, and thus become square-integrable and easy to normalize. In practice, this means that QNMs can be easily normalized by integrating over the whole numerical space including the PML regions [Sau13].

PMLs are very convenient, but they are not necessarily required. Another approach to normalize QNMs relies on the fact that QNMs are poles of the scattering operator. With any modern frequency-domain Maxwell's equation software, the scattered field can be computed for complex frequencies very close to the pole. Since for $\omega \approx \widetilde{\omega}_m$, the scattered field of any resonator is proportional to the normalized QNM field $\widetilde{\mathbf{E}}_m(\mathbf{r})$ with an analytically known proportionality factor [Bai13,Lal18], the *normalized* mode can be obtained just by computing scattered fields at complex frequencies with or without PMLs.

Periodic structures deserve special care. The normalization relies on two QNMs with opposite Bloch wavevectors [Lec07]. For illustration, let us consider a grating that is periodic in the $x$-direction, translationally symmetric in the $z$-direction, and aperiodic in the $y$-direction. Denoting by $k_x$ the $x$-component of the Bloch wavevector, we may write the QNM field as $[\widetilde{\mathbf{E}}_{k_x}(\mathbf{r}), \widetilde{\mathbf{H}}_{k_x}(\mathbf{r})] = [\widetilde{\mathbf{e}}_{k_x}(\mathbf{r}), \widetilde{\mathbf{h}}_{k_x}(\mathbf{r})] \exp(ik_x x + ik_y y)$, $\widetilde{\mathbf{e}}_{k_x}$ and $\widetilde{\mathbf{h}}_{k_x}$ being periodic. If the constitutive materials are reciprocal, as it is assumed in the benchmarks, there always exists a QNM with an opposite propagation constant $-k_x$[Lec07], $[\widetilde{\mathbf{E}}_{-k_x}(\mathbf{r}), \widetilde{\mathbf{H}}_{-k_x}(\mathbf{r})] = [\widetilde{\mathbf{e}}_{-k_x}(\mathbf{r}), \widetilde{\mathbf{h}}_{-k_x}(\mathbf{r})] \exp(-ik_x x + ik_y y)$, and the normalization becomes

$$\iiint_{\Omega} \left[ \widetilde{\mathbf{E}}_{k_x}(\mathbf{r}) \cdot \frac{\partial(\omega\boldsymbol{\varepsilon})}{\partial\omega} \widetilde{\mathbf{E}}_{-k_x} - \widetilde{\mathbf{H}}_{k_x} \cdot \frac{\partial(\omega\boldsymbol{\mu})}{\partial\omega} \widetilde{\mathbf{H}}_{-k_x} \right] d^3\mathbf{r} = 1, \tag{7}$$

where the integral runs over the unit cell for periodic crystals (example 1) or a unit cell including PMLs for gratings (example 2). Note that $(\widetilde{\mathbf{E}}_{-k_x}, \widetilde{\mathbf{H}}_{-k_x})$ can be deduced from $(\widetilde{\mathbf{E}}_{k_x}, \widetilde{\mathbf{H}}_{k_x})$ for symmetric gratings. For instance, in the grating example 2, $\varepsilon(x) = \varepsilon(-x)$ and we have $\hat{\mathbf{x}} \cdot \widetilde{\mathbf{E}}_{-k_x}(-x, y) = \hat{\mathbf{x}} \cdot \widetilde{\mathbf{E}}_{k_x}(x, y)$, $\hat{\mathbf{y}} \cdot \widetilde{\mathbf{E}}_{-k_x}(-x, y) = -\hat{\mathbf{y}} \cdot \widetilde{\mathbf{E}}_{k_x}(x, y)$ and $\hat{\mathbf{z}} \cdot \widetilde{\mathbf{H}}_{-k_x}(-x, y) = \hat{\mathbf{z}} \cdot \widetilde{\mathbf{H}}_{k_x}(x, y)$. Otherwise, two QNMs have to be computed, one with $k_x$ and the other one with $-k_x$ [Lec07]. Figure 1 illustrates the symmetry and the absence of symmetry between $(\widetilde{\mathbf{E}}_{-k_x}, \widetilde{\mathbf{H}}_{-k_x})$ and $(\widetilde{\mathbf{E}}_{k_x}, \widetilde{\mathbf{H}}_{k_x})$ for symmetric and non-symmetric gratings, respectively. Note that in the non-symmetric case, two different calculations will in general yield two different values for the normalized field $\widetilde{\mathbf{E}}_{k_x}$, but the product $\widetilde{\mathbf{E}}_{k_x} \cdot \widetilde{\mathbf{E}}_{-k_x}$ is uniquely defined [Wei17].

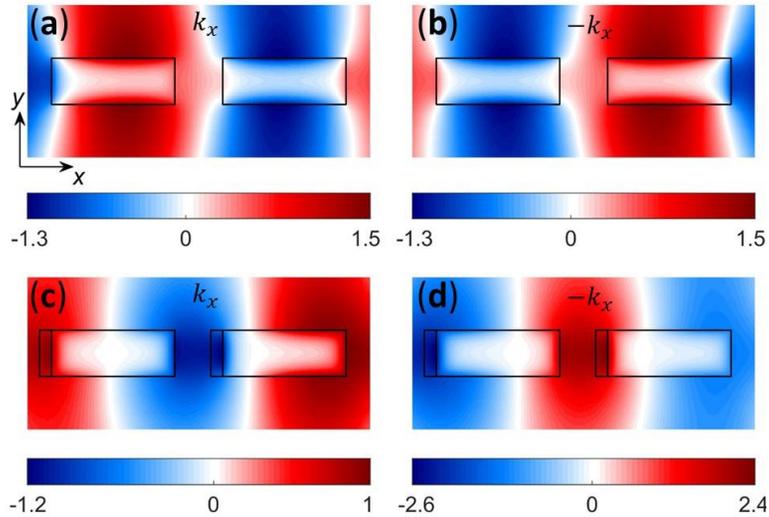

**Figure 1.** Normalized magnetic field distributions for $k_x$ and $-k_x$ and for a symmetric grating (**a-b**), $\varepsilon(x) = \varepsilon(-x)$, and a non-symmetric one (**c-d**), $\varepsilon(x) \neq \varepsilon(-x)$. The symmetric case corresponds to Example 2 (see Section 3.2 for any details on the geometry, polarization ...). Note that $\hat{\mathbf{z}} \cdot \widetilde{\mathbf{H}}_{-k_x}(x, y) = \hat{\mathbf{z}} \cdot \widetilde{\mathbf{H}}_{k_x}(-x, y)$. A dielectric inclusion ($\varepsilon = 2.25$) has been added to break the symmetry in (**c-d**). In that case, there is no symmetry relation between $\widetilde{\mathbf{H}}_{k_x}$ and $\widetilde{\mathbf{H}}_{-k_x}$. The normalized field maps are computed with the FMM3. Note that the colorbars in (**c**) and (**d**) are different.

## 2.3 Mode volume

Just like the quality factor in the temporal domain, the mode volume is an important parameter of resonators in the spatial domain [Rig97]. Initially used by the cavity quantum electrodynamics community for high-$Q$ microcavities, the mode volume was considered as a *real* quantity, $V(\boldsymbol{r}_0) = \frac{\iiint \left[\varepsilon|\tilde{E}|^2 + \mu_0|\tilde{H}|^2\right] d^3 r}{2\varepsilon(\boldsymbol{r}_0)|\tilde{E}(\boldsymbol{r}_0)\cdot \mathbf{u}|^2}$, which gauges the coupling strength of an emitting dipole located at $\boldsymbol{r}_0$ and parallel to the direction $\mathbf{u}$ ($|\mathbf{u}| = 1$) with the cavity mode [Lal18,Ger03]. The mode volume is usually defined for dipoles placed at the field-intensity maximum, where the coupling is also maximum, but it may also be considered as spatially-dependent to directly take into account the dependence of the coupling strength with the dipole position. As one escapes from the Hermitian limit $Q \to \infty$, the definition of a mode volume through energy consideration has obvious shortcomings and is largely inaccurate for low-$Q$ resonators [Lal18]. The real mode volume $V$ should then be replaced by a complex one $\tilde{V} = \frac{\iiint \left(\frac{\partial \omega \varepsilon}{\partial \omega}\tilde{E}^2 - \mu_0\tilde{H}^2\right) d^3 r}{2\varepsilon(\boldsymbol{r}_0)\left(\tilde{E}(\boldsymbol{r}_0)\cdot \mathbf{u}\right)^2}$, which tends towards the classical one in the limit of $Q \to \infty$. For a normalized mode, $\tilde{V}$ is inversely proportional to the square of the normalized modal electric field at the emitter position $\boldsymbol{r}_0$ and takes the simple expression

$$\tilde{V} = \left[2\varepsilon(\mathbf{r}_0)\left(\tilde{\mathbf{E}}(\mathbf{r}_0)\cdot \mathbf{u}\right)^2\right]^{-1}. \tag{8}$$

Complex $\tilde{V}$s are rooted in important phenomena of light-matter interactions in non-Hermitian open systems [Lal18]. For instance, the ratio $\mathrm{Im}\,\tilde{V}^{-1}/\mathrm{Re}\,\tilde{V}^{-1}$ quantifies the spectral asymmetry of the mode contribution to the modification of the Lorentzian shape of the spontaneous-emission-rate spectrum (Purcell effect) of an emitter weakly coupled to a cavity [Sau13]. For strong coupling, it modifies the usual expression of the Rabi frequency by blurring and moving the boundary between the weak and strong coupling regimes [Lal18,Las18], and for cavity perturbation theory, the ratio $\mathrm{Im}\,\tilde{V}^{-1}/\mathrm{Re}\,\tilde{V}^{-1}$ also directly impacts the narrowing or broadening of the resonance linewidth due to the perturber [Lal18,Cog18]. The third 3D example benchmarks $\tilde{V}$ for a plasmonic antenna.

## 2.4 Overview of QNM solvers

For resonators with complex shapes and materials, analytic solutions for QNMs are not available and the continuous Maxwell's operator has to be approximated using a numerical discretization scheme, which preserves the outgoing-wave condition at $|\boldsymbol{r}| \to \infty$. It is not the purpose of this Section to review all the methods that can be used to compute QNMs. The literature is so vast that we restrict the discussion to methods that have been used, not only to compute, but also to normalize QNMs.

In practice, the numerical discretization of the Maxwell's operator is correctly implemented only for a finite spectral interval. Thus only a sub-set of the true QNMs is accurately recovered, namely the states for which the outgoing-wave conditions and discretization are well implemented, and other discrete numerical eigenmodes, the so-called PML modes [Vi14a,Vi14b,Yan18], emerge as a direct consequence of the truncation of the open space. Unlike QNMs, PML-modes depend on the PML material and geometric parameters.

QNMs are preferentially computed by solving Eq. (1) with frequency domain methods operating at a complex frequency. A first option is to calculate the QNMs from a Fredholm type integral equation [Las13,Ber16,Pow17,Pow14, Zhe14,Kri12], in which case the outgoing-wave condition is perfectly fulfilled by construction. However, since the QNM resonance frequency (i.e., the unknown) enters in the outgoing-wave condition, the integral equation defines a nonlinear problem even for nondispersive materials, thereby requiring particular care.

An alternative way is to surround the resonator by PMLs. For nondispersive materials with frequency-independent permittivities and permeabilities, Eq. (1) defines a linear eigenvalue problem and various mode solvers, including commercial ones such as COMSOL Multiphysics [COMSOL], are available to compute many QNMs very efficiently. Normalization is then performed by evaluating the integral of Eq. (6) (or Eq. (7) in the case of periodic structures) over the whole numerical space including the PML regions. For the general case of resonators made of *dispersive materials*, for which Eq. (1) defines a *nonlinear eigenvalue problem*, one needs to know the analytic continuation of the permittivity and permeability tensors, $\boldsymbol{\varepsilon}(\omega)$ and $\boldsymbol{\mu}(\omega)$, at complex frequencies. This strict requirement is usually met by using physical models, which provide fully-analytic expression for the material parameters. This is for instance the case for the Drude electrical conduction for free carriers in metals or highly-doped semiconductors. Alternatively, one may fit material parameters measured at real frequencies to ad hoc analytic expressions, such as multiple-pole Lorentz-Drude expansions [Zha13,Pow17,Ram10], which guarantee causality and the frequency symmetry ($\varepsilon(-\omega^*) = \varepsilon^*(\omega)$) resulting from the real nature of the susceptibility. Systematic and effective procedures for fitting experimental data exist, see for instance the procedure developed with Hermitian functions in the form of polynomial fractions and proposed in [Gar17].

Three general approaches to compute and to normalize QNMs are used in general.

**Pole-search approach**

The pole-search approach is probably the simplest and most general method. It relies on the fact that the resonator response to any driving field diverges as the driving frequency approaches a QNM eigenfrequency. Thus the QNMs can be computed by searching poles in the complex frequency plane for some representative quantities that are complex functions of the frequency, such as the electromagnetic field response to a driving source or some elements of the discretized scattering matrix [Pow14,Zhe14]. Some iterative algorithms, such as the Newton method, are well suitable for the pole searching [Kra00]. They usually require an initial guess value as close as possible to the actual QNM eigenfrequency for fast numerical convergence, and compute QNMs one by one by iteratively exploring the complex plane around every QNM pole. Pole-search approaches are particularly relevant when only a few QNMs need to be computed.

Alternatively, a non-iterative method, the so-called Cauchy Integration Method, has also been developed [Zol05,Del67]. This method is capable to find all the poles in a closed predetermined region of the complex frequency plane, but it needs an extra computational cost associated with the contour integration over the outer boundary of the closed region to invert the discretized Maxwell's matrix [Byk13,Pow14]. For a better accuracy, the pole found with the Cauchy Integration Method can be further refined with the Newton's method [Pow14].

A pole-search QNM-solver freeware, QNMPole, which computes and normalizes QNMs for arbitrary resonators by approximating the inverse of the electric field response with a Padé approximant is available since 2013 [Bai13]. The freeware can be used with any Maxwell's equations solver, including commercial software such as COMSOL Multiphysics.

**Auxiliary-field eigenvalue formulation**

A different approach consists in computing all the QNMs "at one time" by solving a linearized version of the eigenvalue problem, for which a myriad of efficient and stable numerical methods exists. A general approach consists in transforming the nonlinear eigenvalue problem into a linear one by introducing auxiliary fields to account for material dispersion. Several variants exist, but for the sake of simplicity, we will just provide a generic presentation of auxiliary-field techniques [Ram10,Zha13,Luo10,Che06,Taf13,Zi16a,Zi16b,Bru16,Yan18]. The latter have been initially used for computing band diagrams of dispersive photonic crystals [Ram10,Zha13,Luo10,Che06,Bru16] and in time-domain for modelling wave propagation in dispersive media [Taf13,Yan18].

For the sake of illustration, we consider an isotropic (to simplify) medium with a dispersive permittivity described by the single-pole Lorentz model, $\varepsilon(\omega) = \varepsilon_0 \varepsilon_\infty \left( 1 - \frac{\omega_p^2}{\omega^2 - \omega_0^2 + i\omega\gamma} \right)$, and a nondispersive permeability $\mu = \mu_0$. We introduce two auxiliary fields, the polarization $\mathbf{P} = -\varepsilon_0 \varepsilon_\infty \frac{\omega_p^2}{\omega^2 - \omega_0^2 + i\omega\gamma} \mathbf{E}$ and the current density $\mathbf{J} = -i\omega\mathbf{P}$. With elementary algebraic manipulations, we can reformulate Eq. (1) into an extended eigenvalue problem

$$\begin{bmatrix} 0 & -i\mu_0^{-1}\nabla\times & 0 & 0 \\ i(\varepsilon_0\varepsilon_\infty)^{-1}\nabla\times & 0 & 0 & -i(\varepsilon_0\varepsilon_\infty)^{-1} \\ 0 & 0 & 0 & i \\ 0 & i\omega_p^2\varepsilon_0\varepsilon_\infty & -i\omega_0^2 & -i\gamma \end{bmatrix} \begin{bmatrix} \tilde{\mathbf{H}}_m(\mathbf{r}) \\ \tilde{\mathbf{E}}_m(\mathbf{r}) \\ \tilde{\mathbf{P}}_m(\mathbf{r}) \\ \tilde{\mathbf{J}}_m(\mathbf{r}) \end{bmatrix} = \widetilde{\omega}_m \begin{bmatrix} \tilde{\mathbf{H}}_m(\mathbf{r}) \\ \tilde{\mathbf{E}}_m(\mathbf{r}) \\ \tilde{\mathbf{P}}_m(\mathbf{r}) \\ \tilde{\mathbf{J}}_m(\mathbf{r}) \end{bmatrix}, \tag{9}$$

The approach can be straightforwardly extended to multiple-pole Lorentz models by increasing the number of auxiliary fields. It is worth mentioning that the support of both auxiliary fields $\mathbf{P}$ and $\mathbf{J}$ is not necessary in the whole computational domain but only in the subdomains that contain the dispersive material. Note that the Drude model, a particular case for which $\omega_0 = 0$, requires a single auxiliary field $\mathbf{J}$.

QNM eigensolvers based on an auxiliary-fields method have been initially implemented with finite-difference methods [Zi16a,Zi16b]. The latter may introduce inaccuracies for complex geometries, which may lead to the prediction of spurious modes when discretizing curved metallic surfaces on a rectangular grid for instance. QNM solvers based on finite element methods may be preferable. In [Yan18], a general freeware using the COMSOL Multiphysics platform is tested.

**Polynomial eigenvalue formulation**

A purely algebraic linearization is possible, in which the auxiliary fields are no longer related to physical quantities such as the polarization vector and the current density, but to successive time derivatives of the electric or magnetic field. Consider for instance a non-magnetic dispersive medium with a dielectric permittivity described by the same single-pole Lorentz model as in the previous paragraph, $\varepsilon_r(\omega) = \varepsilon_\infty \left( 1 - \frac{\omega_p^2}{\omega^2 - \omega_0^2 + i\omega\gamma} \right)$. The source free propagation equation for the electric field $c^2 \nabla \times \nabla \times \mathbf{E} + \omega^2 \varepsilon_r(\omega)\mathbf{E} = 0$ can be written as a polynomial eigenvalue problem $\omega^4 \mathcal{M}_4 \mathbf{E} + \omega^3 \mathcal{M}_3 \mathbf{E} + \omega^2 \mathcal{M}_2 \mathbf{E} + \omega \mathcal{M}_1 \mathbf{E} + \mathcal{M}_0 \mathbf{E} = 0$, where the operators $\mathcal{M}_i$ are given by $\mathcal{M}_4 = \varepsilon_\infty$, $\mathcal{M}_3 = i\gamma\varepsilon_\infty$, $\mathcal{M}_2 = c^2 \nabla \times \nabla \times (\cdot) - \varepsilon_\infty(\omega_p^2 + \omega_0^2)$, $\mathcal{M}_1 = i\gamma c^2 \nabla \times \nabla \times (\cdot)$ and $\mathcal{M}_0 = -\omega_0^2 c^2 \nabla \times \nabla \times (\cdot)$. Introducing three extra fields $\mathbf{E}_1 = \omega\mathbf{E}$, $\mathbf{E}_2 = \omega\mathbf{E}_1$, and $\mathbf{E}_3 = \omega\mathbf{E}_2$, a generalized linear eigenvalue problem of the form $\mathbf{LV} = \omega\mathbf{RV}$ is obtained:

$$\begin{bmatrix} -\mathcal{M}_0 & -\mathcal{M}_1 & -\mathcal{M}_2 & -\mathcal{M}_3 \\ 0 & 1 & 0 & 0 \\ 0 & 0 & 1 & 0 \\ 0 & 0 & 0 & 1 \end{bmatrix} \begin{bmatrix} \mathbf{E} \\ \mathbf{E}_1 \\ \mathbf{E}_2 \\ \mathbf{E}_3 \end{bmatrix} = \omega \begin{bmatrix} 0 & 0 & 0 & \mathcal{M}_4 \\ 1 & 0 & 0 & 0 \\ 0 & 1 & 0 & 0 \\ 0 & 0 & 1 & 0 \end{bmatrix} \begin{bmatrix} \mathbf{E} \\ \mathbf{E}_1 \\ \mathbf{E}_2 \\ \mathbf{E}_3 \end{bmatrix}. \tag{10}$$

As opposed to the physically-meaningful auxiliary fields $\mathbf{P}$ and $\mathbf{J}$, the supplementary fields $\mathbf{E}_1$, $\mathbf{E}_2$, and $\mathbf{E}_3$ have the same support as the electric field $\mathbf{E}$, i.e., the whole computational domain. Hence, both linearization schemes clearly lead to distinct discrete systems.

# 3. Numerical results and comparison

Several groups in Europe and in China have been contacted to participate in the benchmark using their in-house developed software. In the following sections, three different numerical exercises are benchmarked in relation with the computation and normalization of QNMs.

In Table 1 we present the twelve different implementations of fully-vectorial methods that have been benchmarked. We have classified these implementations into three general categories: modal methods (MM), finite- difference (FD) methods, finite-element methods (FEM). The methods represent a selection of the most popular numerical methods used nowadays in computational electrodynamics.

Clearly, they do not cover all existing methods, but nevertheless the results of the benchmark may be used with confidence by other researchers in the field to test their in-house or commercial software.

| Numerical method | Affiliation | Completed Benchmark | Acronym |
|---|---|---|---|
| Finite difference | Delft-TU | 3 | FD |
| Finite element | ZIB, Berlin | 1-3 | FEM1 |
| a-Fourier Modal Method | Nankai Univ. | 1-3 | FMM1 |
| Finite element (COMSOL) | LP2N, Bordeaux | 1-3 | FEM3 |
| a-Fourier Modal Method | LCF, Palaiseau | 1-3 | FMM3 |
| Finite element | | 1-3 | FEM2 |
| Fourier Modal Method | Stuttgart Univ. | 1-2 | FMM2 |
| Finite element | Institut Fresnel, Marseille | 1-3 | FEM4 |

**Table 1**. Benchmarked methods. The last column presents the acronyms used throughout the article. All methods are performed with in-house developed software, except FEM1 and FEM3 that rely on JCMSuite and COMSOL Multiphysics.

## 3.1 Example 1: plasmonic crystals

We consider as the first example the case of a two-dimensional (2D) plasmonic crystal composed of a periodic array of metallic squares in air. The structure is depicted in Fig. 2(a). It was previously studied by Raman *et al.* with a finite difference frequency domain method [Ram10], and then by Brûlé *et al.* with a finite elements method [Bru16]. In both studies, the dispersion diagram and modes of this absorbing and dispersive system have been computed by implementing an auxiliary-field formulation with a finite difference frequency domain method, but different results have been reported. The period of the square lattice is denoted by $a$ and the size of the square metallic inclusions is $w = 0.25a$. The relative permittivity of the metal is described by a Drude model of the form $\varepsilon(\omega) = 1 - \omega_p^2/(\omega^2 + i\omega\gamma)$, with $\omega_p a/(2\pi c) = 1$ and $\gamma = 0.01\omega_p$.

We first present the dispersion diagram of the plasmonic crystal (complex eigenfrequencies as a function of the wavevector) and then we compare the results given by seven different numerical methods. The benchmark consists of the calculation of one mode (complex eigenfrequency and normalized field) for a given value of the wavevector.

### 3.1.1 Dispersion diagram

We have computed the QNMs of the plasmonic crystal in TE polarization (magnetic field along the $z$ direction) for different values of the wavevector $k_x$ and $k_y = 0$. Figure 2(b) displays the real part of the eigenfrequencies $\widetilde{\omega}_m$ as a function of the wavevector $k_x$ and Fig. 2(c) shows the distribution of the eigenfrequencies in the complex frequency plane. The calculations have been performed with the FEM2.

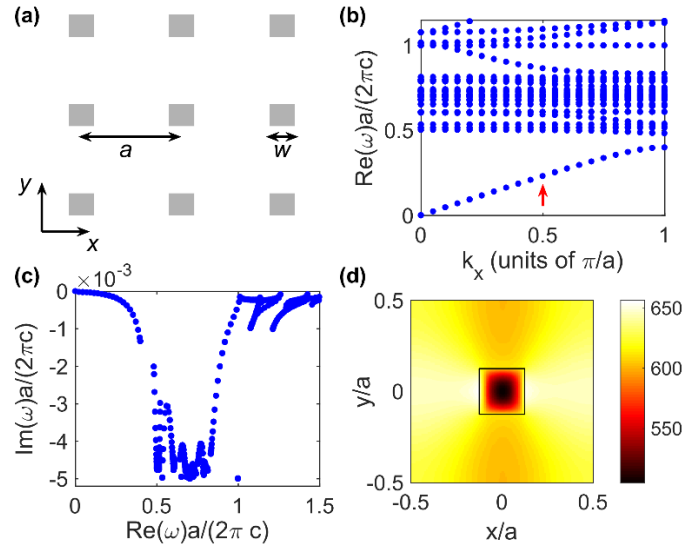

**Figure 2.** Band diagram of a 2D plasmonic crystal. (**a**) The structure consists of a 2D array of metallic inclusions in air. The size of the square Drude inclusions is $w = 0.25a$. (**b**) Real part of the QNMs eigenfrequencies as a function of the wavevector $k_x$ for $k_y = 0$. (**c**) Distribution of the eigenfrequencies in the complex frequency plane for the same wavevector values. (**d**) $a|H_z|$ distribution for the normalized QNM with the lowest frequency for $k_x = 0.5\pi/a$ (red arrow in (**b**)).

A remarkable feature of the dispersion diagram is the existence of many flat bands. The latter correspond to slow high-order surface-plasmon QNMs [Yan18] with high parallel momenta and a strong degree of confinement at the metal/air interfaces. They are highly sensitive to the radius of curvature of the wire corners as we have shown by several computations performed by slightly rounding the corners, in particular in the spectral interval corresponding to $-1/3 < \mathrm{Re}[\varepsilon(\omega)] < -3$ [Bon14a,Bon14b,Bon18,Dem18].

It is noticeable that the spectral positions of these bands are significantly different from those reported in Fig. 2 in [Ram10], although they are computed exactly for the same geometry. To check the numerical results, we have computed the dispersion diagram with three other numerical methods, FEM3 and FEM4 and one in-house FD method specifically implemented at LCF for this benchmark. An excellent agreement between the four numerical methods has been obtained, providing virtually the same spectra for the flat bands. Since the four methods have been implemented independently, we are inclined to think that the results reported in [Ram10] suffer from numerical inaccuracies.

### 3.1.2 Numerical benchmark for a single mode

We have benchmarked several methods for computing the eigenfrequency and the normalized field of the QNM with the lowest frequency at $k_x = 0.5\pi/a$, red arrow in Fig. 2(b). Its magnetic-field modulus is shown in Fig. 2(d). The field is calculated at the center of the metallic inclusion and normalized according to Eq. (7). Table 2 summarizes the most accurate results obtained by each method.

To improve the numerical accuracy, the three FMM implementations increase the number of Fourier harmonics $2N + 1$ retained in the computation. The four FEM implementations rely on different strategies. FEM1 increases the FE order $p$ from 1 to 6, whereas FEM2-4 increase the number $M$ of mesh elements for a fixed FE order. FEM2 and

FEM4 are based on a second order FE and FEM3 is using a fourth order. These results correspond to the data obtained with the highest number of Fourier harmonics, the largest FEM order or the finest mesh, except for FMM3, for which the data in Table 2 are obtained by an extrapolation of the convergence curve, see Section 4.4.

| Method | $\mathrm{Re}(\widetilde{\omega})a/(2\pi c)$ | $\mathrm{Im}(\widetilde{\omega})a/(2\pi c)$ | $\mathrm{Re}(\widetilde{H}_z)a$ | $\mathrm{Im}(\widetilde{H}_z)a$ | Refinement parameter |
|--------|------------|------------|-----------|-----------|----------------------|
| FMM1 | **0.23107**438 | **-0.0001445**244 | **3.32**867 | **-505.0**74 | $N = 100$ |
| FMM2 | **0.23107**368 | **-0.0001440**116 | **3.330**18 | **-505.06**5 | $N = 50$ |
| FMM3 | **0.23107**371 | **-0.0001440087** | **3.330**23 | **-505.06**7 | Extrapolation |
| FEM1 | **0.23107**370 | **-0.0001440083** | **3.329**96 | **-505.06**2 | $p = 6$, $M =$ 1133 |
| FEM2 | **0.23107**047 | **-0.0001441**258 | **3.329**00 | **-505.**130 | $p = 2$, $M = $ 40k |
| FEM3 | **0.23107**349 | **-0.0001440**188 | **3.329**91 | **-505.06**1 | $p = 4$, $M = $ 66k |
| FEM4 | **0.23107**496 | **-0.0001439**568 | **3.32**819 | **-505.0**41 | $p = 2$, $M = $ 55k |

**Table 2.** Most accurate numerical values obtained for the complex eigenfrequency and the normalized magnetic field $\widetilde{H}_z a$. The last column summarizes the values of the refinement parameter used to obtain the data. Note that $\widetilde{H}_z a$ is not a dimensionless quantity and is expressed in $\mathrm{A.s.m^{-1/2}.kg^{-1/2}}$. For FMMs, $N$ represents the truncation rank, and for FEMs, $p$ is the FE order and $M$ is the number of mesh elements. Bold digits (in the present Table and in following ones) are believed to be correct.

Figure 3 shows how the different numerical methods converge towards their best values. The convergence is displayed as a function of the CPU time (or wall time) in seconds. We have represented the relative difference between the data and a reference. For the reference, we chose the mean value of the results given by the three FMM1,2,3 that provide the largest number of common digits in Table 2. Figure 3 evidences that all seven methods converge towards the same values both for the eigenfrequency and for the normalized field. Remarkably, the real and imaginary parts of the eigenfrequency are obtained with at least six significant digits, see the bold numbers in Table 2. The real and imaginary parts of the field are obtained with two and four significant digits, respectively.

For this 2D problem, the wall times required to reach such an excellent accuracy span between a few seconds and one hundred for the slowest. The three FMM implementations converge faster than the FEM implementations. Note that the FEM is generally fast at extracting a large number of the eigenvalues spectrum at once, whereas the benchmark concerns a single eigenvalue. Note also that increasing the FEM order instead of refining the mesh leads to a faster convergence.

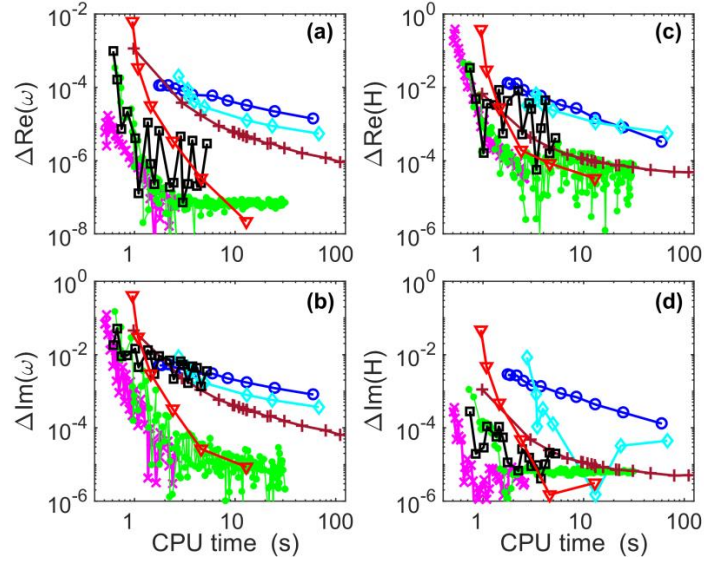

**Figure 3.** Convergence performance obtained for the QNM with the lowest frequency at $k_x = 0.5\pi/a$. Data are displayed as a function of the CPU time in seconds. We present the relative difference between every calculation and a reference value obtained by averaging the results of FMM2, FMM3, and FEM1. (**a**) Real part of the eigenfrequency. (**b**) Imaginary part. (**c**) Real part of the normalized magnetic field. (**d**) Imaginary part. Black squares: FMM1. Magenta x-marks: FMM2. Green dots: FMM3. Red triangles: FEM1. Blue circles: FEM2. Brown pluses: FEM3. Cyan diamonds: FEM4.

## 3.2 Example 2: metal grating

### 3.2.1 General overview of the structure

The second benchmark considers an open structure: a 1D grating, made of a periodic slit array etched in a free-standing gold membrane, see Fig. 4(a). The geometry is based on [Col10]. The period is $a = 482.5$ nm, the rod width $w = 347.5$ nm and the rod height $h = 130$ nm. A Drude gold relative permittivity is considered, $\varepsilon_r(\omega) = 1 - \omega_p^2/(\omega^2 + i\gamma\omega)$ with $\omega_p = 1.15\text{e}16 \ s^{-1}$ and $\gamma = 0.0078 \ \omega_p$. This model agrees well with experimental ellipsometric data in the near infrared, but fails to describe the response of gold in the blue region of the spectrum.

The grating exhibits a valuable bandpass filtering behavior in transmission [Col10]. The specular transmission $T_0$ and absorption A are shown in Figs. 4(b) and (c) for 3 in-plane wavevector $k_x$ values and for TM polarization (*i.e.* $H_z$ component parallel to the wires). The spectra are obtained using a collection of direct FE computations by iterating over the wavelength and the parallel momentum of the incoming plane wave [Dem07]. Note that in the frequency domain, handling the frequency dispersion of materials is trivial since the permittivity can be updated from tables as the frequency is swept. The goal of the second benchmark is to retrieve with accuracy the QNMs accounting for the main resonant transmission peaks in Figs. 4(b) and 4(c).

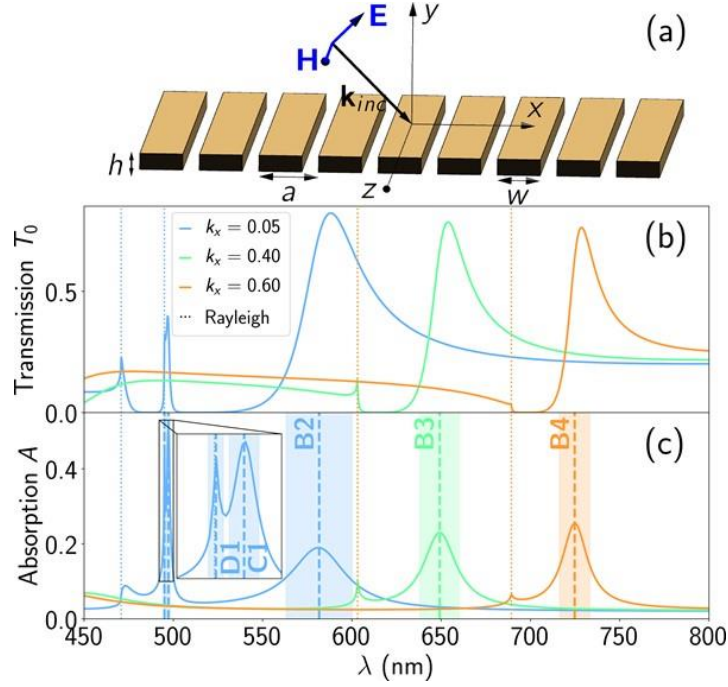

**Figure 4 (a)** Geometry of the free-standing 1D grating. **(b)** Specular transmission $T_0$ for three values of $k_x$, $k_x a/\pi = 0.05, 0.4, 0.6$. **(c)** Absorption $A$ (fraction of the incident energy absorbed) spectra. The dotted vertical lines represent the Rayleigh anomalies. The dashed lines represent the real parts of the grating QNM eigenfrequencies $2\pi c/\tilde{\omega}$ taken from Fig. 5. The widths of the rectangular patches centered around each dashed line are set to be $2\mathrm{Im}(2\pi c/\tilde{\omega})$, encoding the quality factor of each resonance. The inset is a zoom with the same color conventions, showing near normal incidence sharp resonances and the corresponding QNMs labelled C1 and D1 in Fig. 5.

### 3.2.2 Dispersion relation

The dispersion relation of the grating is shown in the right panel of Fig. 5. It has been computed using FEM4 (crosses) and a pole search method based on FMM3 (small dots forming a seemingly continuous line). The real parts of the normalized eigenvalues $\tilde{\omega}/\eta$ with $\eta = 2\pi c/a$ are represented in ordinate. The abscissa represents the normalized Bloch wavevector in the first reduced Brillouin zone. The imaginary part of the normalized eigenvalues is encoded into a jet color scale of the dots and crosses that form the dispersion curves. For $\mathrm{Re}(\tilde{\omega})/\eta < 1.2$, the dispersion relation exhibits six branches, labeled "A", "B" … "F" and standing below the folded light line shown with dashed gray lines. The insets show the QNM fields $|\tilde{\mathbf{H}}|$.

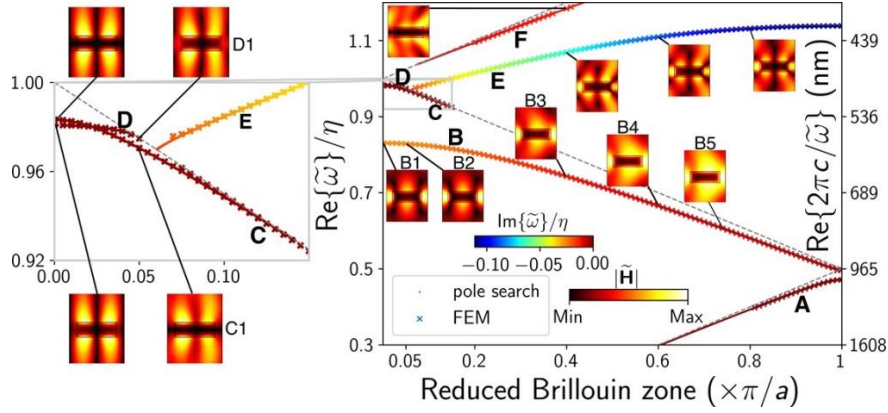

**Figure 5** Dispersion relation of the gold grating. Crosses represent the real parts of the eigenvalues obtained with FEM4. Dots forming an almost continuous curve represent the real parts of the eigenvalues found using a complex pole search. The insets represent the modulus of the magnetic field distribution of the QNM corresponding to the eigenvalue indicated by a black line. The left panel shows a zoom close to the center of the Brillouin zone.

With the help of the dispersion curves, we may interpret the main features of the transmission and absorption spectra of Fig. 4(b)-(c). Let us begin with band B, for which $\mathrm{Re}(\widetilde{\omega})/\eta$ lies between 0.493 and 0.83, corresponding to wavelengths in the interval [581; 979] nm. The QNMs can be directly linked to the absorption spectrum, when superimposing to the absorption peaks of Fig. 4(c) with rectangular patches centered at $\mathrm{Re}(2\pi c/\widetilde{\omega})$ with widths $2\mathrm{Im}(2\pi c/\widetilde{\omega})$. In particular, the real parts shown with dashed vertical lines exactly match the maxima of the absorption spectrum for the three values of $k_x$.

Now, let us move to higher frequencies with the two high sharp absorption peaks ($A \approx 50\%$) observed for $k_x = 0.05\pi/d$ and shown in the zoom Fig. 4(c). Again, when superimposing patches corresponding to $\widetilde{\omega}_{D1}$ and $\widetilde{\omega}_{C1}$, it appears again that these high peaks are attributable to the excitation of the QNMs D1 at 495 nm and C1 at 497 nm (see the zoom in the left panel in Fig. 5). Note that in the inset of Fig. 4(c), the central peak $\mathrm{Re}(2\pi c/\widetilde{\omega}_{D1})$ is almost superimposed with the Rayleigh anomaly corresponding to the first diffraction order. This is consistent with the fact that band D meets the first folded light line for $k_x \gtrsim 0.05\pi/d$ as it can be clearly seen on the zoom of the dispersion relation in Fig. 5. Theses QNMs cannot be excited at normal incidence for symmetry arguments and they cannot be excited either for large $k_x$ values since branches C and D reach the folded light line for $k_x \approx 0.05\pi/d$.

In conclusion, all the resonant features of the absorption spectrum are understood from the dispersion relation. Note that, to recover all the features of the transmission spectra (not only the peaks, but also the zeros of $T_0$ on the blue side of every peak), one needs to compute many QNMs [Gra18].

### 3.2.3 Numerical results for one selected eigenmode

The grating benchmark concerns the numerical computation of a single QNM (labelled B3 in Fig. 5) responsible for the absorption peak centered at 650 nm in Fig. 4(c). Two numerical values are benchmarked: The complex eigenfrequency $\widetilde{\omega}_{B3}a/(2\pi c)$ and the normalized eigenfield at one point $\widetilde{H}_{z,B3}(0,h)$, the origin being chosen at the center of the rod and $h$ being the height of the rod. The results are shown in Table 3 and Fig. 6 for seven different numerical codes. The real and imaginary parts of the normalized eigenfield $a\widetilde{H}_{z,B3}$ are shown in Fig. 6(e-f). The refinement parameter for each method is indicated in the right column. All methods show remarkable agreement, with at least four significant digits for the eigenvalue and three for the eigenfield.

| Method | $\widetilde{\omega}a/(2\pi c)$ | $\widetilde{H}_z$ | Refinement |
|--------|-------------------------------|-------------------|------------|

|        | $\tilde{\omega}_{B3}a/(2\pi c)$ | $\tilde{H}_{z,B3}(0,h)a$ | parameter |
|--------|--------------------------------|--------------------------|-----------|
| FMM1   | **0.74303**−**0.012**662$i$    | **101**.358+**761**.312$i$ | $N = 100$ |
| FMM2   | **0.7430756**−**0.0126**$i$    | **101**.91+**761**.305$i$  | $N = 50$  |
| FMM3   | **0.74307571**−**0.012660590**$i$ | **101**.867+**761**.313$i$ | Extrapolation[a] |
| FEM1   | **0.74307569**−**0.012660593**$i$ | **101**.887+**761**.304$i$ | $p = 6$, $M = 1256$ [b] |
| FEM2   | **0.74305**−**0.0126**644$i$   | **101**.99+**761**.93$i$   | $M = 123$k |
| FEM3   | **0.74304**−**0.0126**608$i$   | **101**.46+**761**.64$i$   | $M = 67$k  |
| FEM4   | **0.7431**0−**0.0126**553$i$   | **101**.89+**760**.79$i$   | $M = 281$k |

**Table 3**. Most accurate numerical values obtained for the complex eigenfrequency $\tilde{\omega}_{B3}a/(2\pi c)$ and the normalized magnetic field $\tilde{H}_{z,B3}(0,h)a$. The refinement parameter for each method is indicated in the upper right column.

[a] The convergence curve is calculated up to $N = 201$ and extrapolated, see Section 4.4.

[b] Refinement by the element polynomial degree $p$, see Section 4.3.

As previously, we have represented in Fig. 6 the relative difference between the data and a reference. For the reference, we chose the mean value of the results given by FMM2,3, and FEM1. These methods are the ones that have the largest number of common digits, see Table 3. The conclusions regarding convergence are very similar to those made for the first benchmark. However, the convergence speed appears to depend significantly on the implementation, with some implementations of the FMM converging slower than the faster implementations of the FEM, see the convergence-rate results of Figs. 6(a)-(d). Among the FEM codes, FEM2-4 have the same refinement parameter, the number $M$ of mesh elements, whereas for the FEM1, the refinement parameters are the element polynomial orders $p$ (up to $p = 6$) and of the mesh refinement around the corners that is predefined with a refinement length $h$ proportional to $10^{-p/2}$.

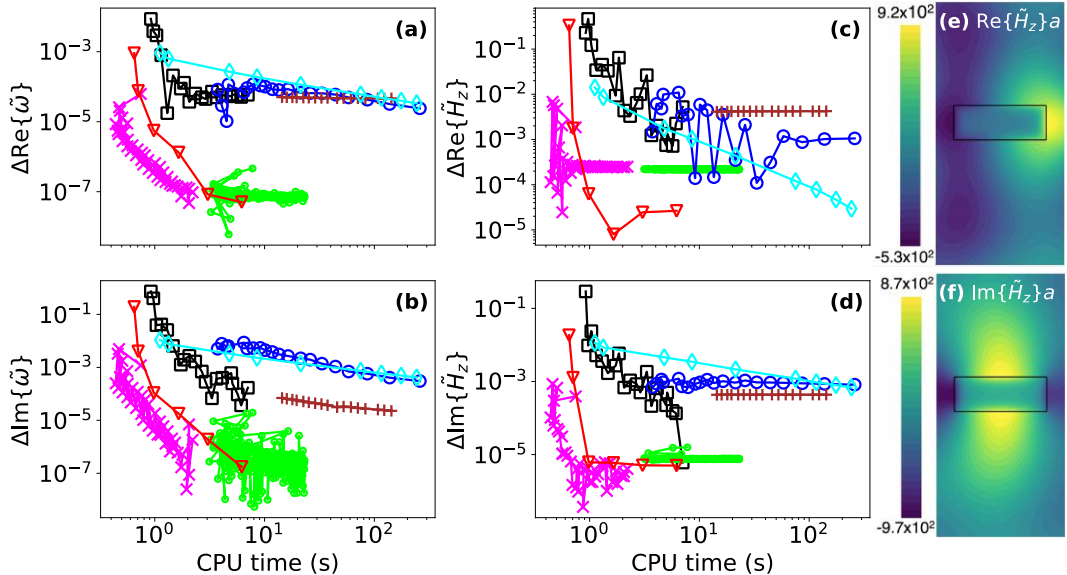

**Figure 6** – Convergence performance obtained for the QNM labelled B3 in Fig. 5 (second lowest eigenfrequency for $k_x = 0.4\pi/a$). Data are displayed as a function of the CPU time. We present the relative difference between every calculation and a reference value obtained by averaging the results of FMM2, FMM3, and FEM1. (**a**) and (**b**) Relative difference for real and imaginary parts of $\tilde{\omega}_{B3}a/(2\pi c)$. (**c**) and (**d**) Relative difference for real and imaginary parts of $\tilde{H}_{z,B3}(0,\text{h})$. The color convention is the same as in Fig. 3: Black squares: FMM1. Magenta x-marks: FMM2. Green dots: FMM3. Red triangles: FEM1. Blue circles: FEM2. Brown pluses: FEM3. Cyan diamonds: FEM4. (**e**) and (**f**) Maps of the normalized QNM $\tilde{H}_{z,B3}$.

## 3.3 Example 3: Nanocube antenna

The third benchmark is related to a 3D geometry, a nanocube antenna that belongs to the family of nanoresonators relying on slow gap-plasmons [Lal18a]. This geometry has recently attracted much attention for various applications and fundamental studies, from biochemical sensors, photodetectors, metamaterials, non-linear switches or light sources [Mor12,Aks14], to the test of spatial dispersion (non-locality) in metals [Cir12]. The geometry is directly inspired by the experimental work in [Aks14], see Fig. 7(a). A 8-nm-thick polymer spacer is sandwiched between two metals, a silver nanocube and a gold substrate. We use a bi-pole Drude model for silver and a quadruple-pole Drude-Lorentz model for gold, to accurately take into account the material dispersion, see details in the caption of Fig. 7(a).

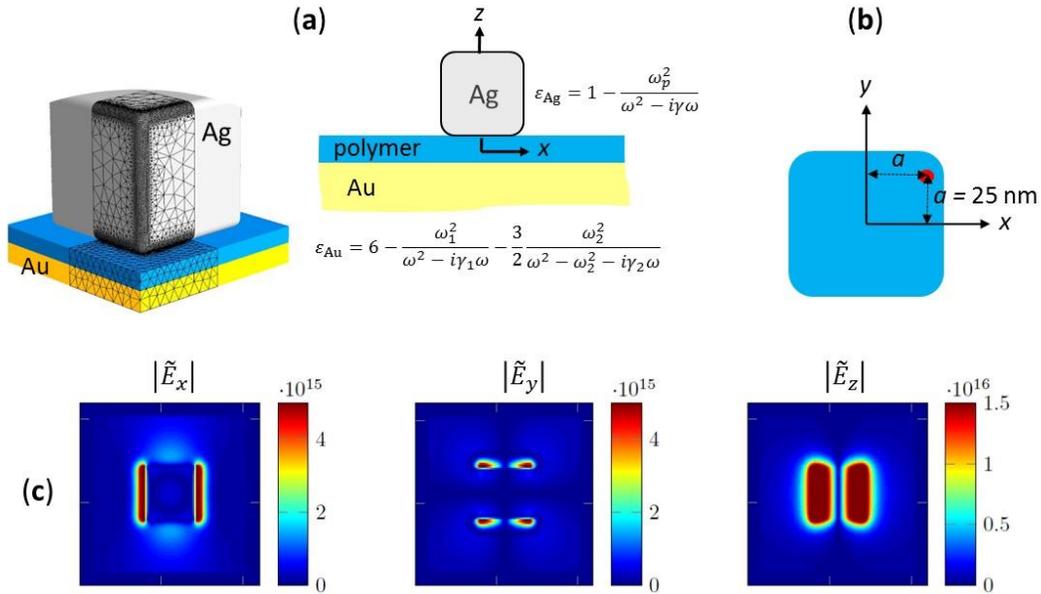

**Figure 7.** Ag nanocube antenna of size 65 nm deposited on a 8-nm-thick polymer film (refractive index 1.5) coated on an Au substrate. (**a**) Silver is modeled as a Drude metal, with $\lambda_p = 2\pi c/\omega_p = 138$ nm and $\gamma = 0.0023\omega_p$. For gold, we use a Drude-Lorentz model, with $\omega_1 = 1.317 \times 10^{16}$ rad/s, $\gamma_1 = 6.216 \times 10^{13}$ rad/s, $\omega_2 = 4.572 \times 10^{15}$ rad/s, and $\gamma_2 = 1.332 \times 10^{15}$ rad/s. The cube corners are rounded with a 4-nm radius. The left panel shows the fine tetrahedral mesh used with FEM1, see Section 4.3. (**b**) The red point, located in the median plane of the polymer ($z = 4$ nm), shows the point $\boldsymbol{r_0}$ where the mode volume is computed. (**c**) Electric field distribution of the fundamental magnetic-dipolar mode that resonates in the red. Owing to the symmetry, another QNM exists at the same complex eigenfrequency with a field distribution rotated by 90° around the $z$-axis. The distributions are computed with the FD method for a 1-nm discretization-grid.

The nanocube geometry offers deep-subwavelength field confinements and enhancements. In the visible spectral range, the response is governed by two dominant QNMs, the fundamental magnetic-dipole-like mode in the red spectrum and an electric-dipole mode in the green spectrum. Only the magnetic mode will be considered hereafter. Its resonance mechanism can be understood from a Fabry-Perot model with a slow gap-plasmon bouncing back and forth between gap extremities, and its resonance linewidth results from several contributions, photon radiation in air, surface-plasmon launching around the nanocube, and metal absorption mainly around the nanogap in the metal. The respective impacts of each contribution on the QNM lifetime have been analyzed comprehensively in [Fag15].

We focus on two quantities related to the fundamental mode, the complex resonance wavelength $\tilde{\lambda}$ and the complex mode volume $\tilde{V}$ of the magnetic-dipole-like mode for the benchmark. Figure 7b shows the precise position $\boldsymbol{r}_0$ of the $z$-polarized dipole source located in the median plane of the polymer film, which is used for the computation of $\tilde{V}$. Figure 7(c) displays the electric-field distribution of the magnetic-dipole mode. The fundamental mode is degenerate, meaning that there are two fields with the same eigenvalue due to the C4v symmetry of the configuration. Any linear combinations of these two modes are again eigenmodes, which renders the selected cases particularly interesting. Only one mode is shown in the Figure, the other mode being obtained by a 90° rotation around the $z$-axis.

Table 4 summarizes the main results obtained for $\tilde{\lambda}$ and $\tilde{V}$ by the five partners. The numbers correspond to the most accurate results obtained for "sampling grids" with the highest resolution. Concerning the resonance wavelength and setting aside the results obtained with FD-TU, an excellent agreement is obtained between the two other methods. Probably due to the rounded corners, the FD scheme suffers from inaccuracies compared to the two FEMs that provide nearly identical results. This explains the 15 nm shift-difference for $\text{Re}(\tilde{\lambda})$ between the second column and others. This shift is rather expected. Because of the tiny modal volume ($Re(\tilde{V}) \approx 4 \times 10^{-5} \mu m^3$), it is inevitable that numerical inaccuracies such as those related to sampling result in large shifts of the resonance frequency (the nanocube is an excellent sensor in the real world, and thus a sensitive testcase for numerics). The five methods rely on the normalization of Eq. (6). They provide very similar mode volumes. Overall, we conclude that a good agreement is achieved; the peak deviations being $< 2\%$ for $\text{Re}\,\tilde{\lambda}$, $\text{Im}\,\tilde{\lambda}$ and $\text{Re}\,\tilde{V}$, and 2.4% for $\text{Im}\,\tilde{V}$.

We conclude that, even for a complicated 3D geometries made of two different metals, a very good accuracy (a few percent for $\tilde{\lambda}$) which is often largely enough for analyzing experimental data can be reached with modest computational efforts (CPU time < 1 min), and this by all methods. We conclude that 3D QNM solvers are available on the shelves.

| Methods | $\tilde{\lambda}\,[\mu m]$ | $\tilde{V}\,[\mu m^3] \times 10^4$ | Refinement type |
|---------|---------------------------|-----------------------------------|-----------------|
| FD | **0.68**10+**0.0159**$i$ | **0.392**+0.030$i$ | 1 nm mesh size |
| FEM1 | **0.665**85+**0.015574**$i$ | **0.384**49+**0.02926**$i$ | $p = 4$, $M = 40$k |
| FEM2 | **0.665**59+**0.015**362$i$ | **0.386**+**0.0283**$i$ | $p = 1$, $M = 276$k |
| FEM3 | **0.665**17+**0.015**468$i$ | **0.384**+**0.0287**$i$ | $p = 2$, $M = 220$k |
| FEM4 | **0.665**01+**0.015**395$i$ | - | $p = 2$, $M = 51$k |

**Table 4.** Nanocube with rounded corners.

Three partners have additionally performed the same computations by considering a perfectly cubic nanocube with sharp corners and edges. The results computed for the highest numerical resolution or extrapolated are show in Table 5. The mean value is $\tilde{\lambda} = 0.6996 + 0.143i$ with a standard deviation $0.0034 + 0.0004i$. A slightly larger deviation is obtained for $\tilde{V}$. Again the agreement is quantitative between all the methods.

| Methods | $\tilde{\lambda}\,[\mu m]$ | $\tilde{V}\,[\mu m^3] \times 10^{-4}$ | Refinement type |
|---------|---------------------------|--------------------------------------|-----------------|
| FMM1 | **0.6974**+**0.01412**$i$ | **0.40**987+**0.02515**$i$ | $N = 55$ |
| FMM3 | **0.697**91+**0.01412**$i$ | **0.41**+**0.025**$i$ | $N = 60$+extrapol. |
| FD | 0.7053+0.0148$i$ | 0.4217+0.0260$i$ | 1 nm mesh size |

**Table 5**. Nanocube with 90° corners.

The benchmark geometry initially proposed was a 3D nanocube with rounded corners, see Fig. 7(a). For simplicity, we have also considered a nanocylinder geometry with exactly the same materials, the cube being just replaced by a silver cylinder with a 70-nm diameter and 65-nm height. For the mode volume, a $z$-polarized electric dipole placed at a distance $25\sqrt{2}$ nm from the cylinder axis is considered. With the axisymmetry, the implementation is easier and more methods have been tested. In addition the computational results are more accurate than for the cube case, as shown by the data in Table 6.

| Methods | $\tilde{\lambda}$ [$\mu m$] | $\tilde{V}$ [$\mu m^3$] $\times 10^{-4}$ | Refinement type |
|---|---|---|---|
| FMM1 | 0.66369+0.01480$i$ | 0. 645+0.0427$i$ | $N = 50$ |
| FMM3 | 0.663557+0.0146866$i$ | 0.63898+0.04616$i$ | $N = 385$ + extrapol. |
| FEM1 | 0.66356+0.014687$i$ | 0.6420+0.045631$i$ | $p = 4$, $M = 1086$ |
| FEM3 | 0.6644+0.01452$i$ | 0.58+0.0400$i$ | $p = 2$, $M = 5k$ |
| FD | 0.6753+0.0150$i$ | 0.523+0.040$i$ | 1 nm mesh size |

**Table 6**. 2D axisymmetry cylinder nanoantenna.

# 4. Method implementation and discussion

This Section provides details that each participant judge important for the readership. It does not include a full presentation of the methods, which have been published elsewhere. However we include all specific modifications made in order to improve the performance.

The FMM (also known as the RCWA) [Moh95,Lal96,Li98] relies on an analytical integration into one (longitudinal) direction of the space and on Fourier series expansions in the two (transversal) others. In the direction of integration, the system is sliced into several layers that have translational symmetry. Maxwell's equations are transformed in each layer into an equivalent algebraic eigenvalue problem, whose eigenmodes propagate or decay within the layers. By using these eigenmodes as a basis for the expansion of an arbitrary field, no discretization is needed in the direction of translational symmetry, making the FMM rather efficient for a restricted numbers of layers. Due to the use of Fourier expansions in the transversal directions, the FMM is restricted to periodic geometries. By incorporating absorbing boundaries or PMLs in the transverse directions, the FMM can be also used for non-periodic structures as well [Sil01]. It is known as the aFMM ("a" for "aperiodic"). The accuracy of the FMM or the aFMM increases with the truncation rank $N$ of the Fourier series, corresponding to a total number of retained Fourier harmonics of $(2N + 1)$ in 2D or $(2N + 1)^2$ in 3D.

The FEM and FD methods are also well established methods that are used in many fields. The interested reader may refer to the textbook [Jin02] for a review in electromagnetism. Both methods rely on a full discretization of space and the accuracy increases with the number of unknowns that is proportional to the number $M$ of mesh elements and increases with the order $p$ of the FE scheme used.

## 4.1 LP2N, Bordeaux

Two freeware are used at LP2N to compute the QNMs of the three benchmarked geometries on a desktop PC with 32

GB RAM and an Intel(R) Xeon(TM) CPU X5660 @ 2.80GHz processor.

The first freeware, *QNMEig*, has been launched in 2018. It includes a QNM-solver [Yan18] to compute and normalize QNMs and PML-modes and, in a second step, to reconstruct the field scattered by the micro and nanoresonators in the QNM basis. In general, accurate reconstructions are obtained with tens or hundreds of QNMs retained in the expansion, depending of the geometry, see details in [Yan18].

QNMEig relies on the eigenfrequency solver of the electromagnetic radio frequency (RF)/optics module of COMSOL Multiphysics [Comsol]. The COMSOL solver computes a large number (set by the user) of QNMs only for non-dispersive media and does not normalize the field. For dispersive media, it approximately transforms nonlinear eigenvalue problems to linear ones by performing second-order Taylor expansion of complex material parameters with respect to some frequency set by user, the so-called linearization point. Thus, such approximate approach cannot accurately find modes away from the linearization point.

QNMEig is in fact an extension of the COMSOL solver that accurately handles dispersive media by coupling the build-in (RF)/optics module and the weak-form module, so that it can be very easily operated by COMSOL users. It relies on the auxiliary-field formulation, see Section 2.4 and [Yan18]. QNMEig normalize the modes by computing the volume integral of Eq. (6). It has been used for the three benchmarks. The CPU-time for solving 3D problems such as the present nanocube antenna of Example 3 or a photonic crystal cavity [Cog18], is typically $\approx 2$ min with a standard desktop computer, see more details in [Yan18]. Two minutes are in general enough to get accurate results for the QNM field and eigenvalue, to challenge experimental data. QNMEig and the companion MATLAB Tooboxes are available at the group webpage [http]. The package includes the four COMSOL model sheets of the benchmarked geometries, the plasmonic crystal, the grating and the nanocube and nanocylinder antennas, and also one tutorial example, a metal sphere in air, for which a document, presenting step by step details on how starting a QNMEig simulation, is offered from the perspective of new users.

The second freeware, QNMPole, is based on the pole-search algorithm described in [Bai13]. It calculates and normalizes the modes of plasmonic or photonic micro/nanoresonators by successive iterations, starting from an initial guess value. The pole-search approach is completely general and can be used with arbitrary geometries and materials, and with a large variety of frequency-domain Maxwell's equations solvers. It has been tested for instance with FMM1, FMM3, FMM2 and FEM3, and could have been used with the other methods as well. For researchers that use COMSOL Multiphysics as a standard Maxwell equation solver operating at real frequencies (electromagnetic RF and optics modules), QNMPole includes a few MATLAB programs that operate under MATLAB-COMSOL livelink and allow the user to compute and normalize QNMs, and customize their specific demands with MATLAB programming. The use of QNMPole is recommended if one just needs to compute a few modes, or if the permittivity of some constitutive materials cannot be cast into a N-pole Lorentz-Drude model (required for QNMEig), or if the material is not reciprocal.

When QNMPole and QNMEig are both used with the same mesh, they provide exactly the same eigenfrequencies and normalized fields. The coincidence has been checked for each benchmark, and for many other examples. This is why only one set of data is shown for the LP2N results. In addition to the QNM-solver, QNMPole and QNMEig also include pedagogical MATLAB toolboxes to reconstruct the scattered field in the QNM bases, to calculate the absorption/extinction cross-sections or the Purcell factor. Finally note the QNMs computed for the grating case are eigenvectors with a complex $\tilde{\omega}$ for a fixed in-plane real wavevector $k_x$, implying that the QNMs are the pole that would be revealed in experiments by varying the frequency of the incident beam, while maintaining for each frequency the product $\omega \sin \theta$ constant. This implies that the angle of incidence $\theta$ be tuned between every spectrum measurement. In reality, however, $\theta$ is usually fixed during the acquisition of the spectrum data. Thus, considering the exact experimental protocol, further investigations into QNM computation and normalization for fixed $\theta$ shall be investigated [Gra18].

## 4.2 TU-Delft

The reduced-order modal solver FD used to determine QNMs in arbitrarily-shaped dispersive 3D structures on open domains is based on a Lanczos-type reduction method that exploits a particular symmetry property of Maxwell's equations. The approach consists in casting the Maxwell equations and the second-order dispersive relations into first-order form using auxiliary field variables and discretize the resulting system in space. Outgoing wave propagation is taken into account via a particular realization of the perfectly matched layer technique in which complex spatial step sizes match the layer PML to the computational domain on a subset of the complex-frequency plane [Dru16,Dru13]. The order of the resulting discretized first-order system is typically very large (millions of unknowns in 3D) so that a direct computation of the QNMs is usually not feasible. Fortunately, the system is symmetric with respect to the discrete counterpart of the bilinear form of Eq. (6) [Lal18] and allows us to efficiently compute the QNMs via a three-term Lanczos-type recursion relation. Only three vectors defined on the total computational domain need to fit into the computational memory due to this three-term relation. Finally, the scattered field of the resonator can be efficiently reconstructed as well without any significant additional computational costs [Zi16a,Zi16b].

The FD solver uses finite-differences to come to a symmetrizable dynamic system which depends nonlinearly on the frequency because of either the dispersive PML used or the dispersion of the material. We stress that FD solver can be used for other types of discretizations that are symmetrizable. In practice, we use a linearized PML that is matched to the frequencies of QNMs of interest, to get rid of the nonlinearity introduced by the PML. Using auxiliary fields, we then obtain a linear shifted system that is symmetric in the bilinear form. The QNMs can then efficiently be computed as the Lanzos-Ritz pairs of this system using short term recurrence relations [Zi16a,Zi16b]. The current implementation of the FD solver relies on a MATLAB prototype implementation with a simple gridding routine for the geometries.

For the 3D nanoantenna geometry, we use second-order finite-differences and consider seven different discretization meshes to study the convergence performance of the FD solver for the geometry with sharp corners and edges with a MATLAB 2016 implementation run on 4 of the 14 cores of a single Intel Xeon E5-2695 v3 CPU at 2.30GHz. The accuracy of the computational results increases quadratically as the discretization step decreases. The computation time behaves linear with the number of degrees of freedom, which themselves increase cubically with the discretization step. Thus, the error expressed in computation time approximately scales as O(time$^{-2/3}$).

Currently the FD method relies on a simple gridding algorithm with straightforward medium averaging to obtain the finite-difference system. Compared to the FEMs, meshing of complex geometries with round corners and edges with finite-difference methods is challenging. We experienced some difficulties while gridding the rounded corners of the nanocube benchmark and therefore only computed the QNMs for the two finest meshes, and our values deviate by approximately 2% from the FEM results in Table 6. Although the error made remains reasonably low for a nanoresonator that is ultrasensitive to fabrication or numerical imperfections, this motivates us to further develop the solver by using more advanced homogenization and averaging schemes to better capture the geometry in the future.

Due to the Lanczos algorithm, 99% of the wall time of the FD-TU method is spent on sparse matrix-vector and vector-vector products. This shows the potential of the method for GPU implementations to obtain faster runtimes. Concerning memory requirements, our algorithm needs to store three double precision complex vectors of the size of the number of degrees of freedom. For a 2-nm discretization, this means that 579 MB of storage is required. In the current implementation the runtime of our algorithm is roughly 1/6000 s/#core/#degree-of-freedom as the runtime scales close to linear with the number of degrees of freedom.

## 4.3 Zuse Institute Berlin

FEM1 relies on the FEM software package *JCMsuite*, which is developed by JCMwave at ZIB and which is also commercially available. *JCMsuite* includes solvers for time-harmonic Maxwell eigenvalue [Zsc06] and scattering [Pom07] problems, as well as for further problem classes. The implementation includes higher-order finite elements, with $h$- and $p$-adaptivity, mesh generators for tetrahedral, prismatoidal and mixed meshes, and adaptive PMLs, which allows to handle also resonance problems with structured exterior domains [Ma13].

For benchmarks 1 and 2, we have used the resonance mode solver included in *JCMsuite* which solves the nonlinear eigenvalue problem using an auxiliary field approach. For computing QNMs with increasing resolution, we use a MATLAB script which defines the physical and numerical project parameters, invokes the solver for QNM computation and for post-processing. In a loop, numerical accuracy is increased stepwise by increasing the polynomial order $p$ of the used finite elements, $p = 1 \dots 6$, and by simultaneously increasing the mesh resolution at the metal corners. The initial guesses for the eigenfrequency is chosen as $\frac{\omega a}{2\pi c} = 0.231$ (Benchmark 1) and $\frac{\omega a}{2\pi c} = (0.74 - 0.013i)$ (Benchmark 2). In each step of the loop, the guess is updated with the prior result. Computation of the normalized field values is performed by computing solutions with Bloch vectors $k_x$ and $k_{-x}$ and applying Eq. (7). Figure 8(a)-(b) shows the relative error of the real and imaginary parts of $\tilde{\omega}$ and $\tilde{V}$. Here, the relative error $\delta A$ of quantity A is defined as deviation from a quasi-exact solution $A_{QE}$, i.e., $\delta A = |A - A_{QE}|/|A_{QE}|$. As quasi-exact solution for the eigenfrequency and normalized field, we use results obtained with the highest numerical accuracy setting, $p = 6$. The relative errors are displayed as function of number of unknowns of the discrete problem which increases with increased finite element degree $p$ and with increased mesh resolution.

For benchmark 3, we have used a recently developed contour integral method based on Riesz projections [Zsc18]. We solve scattering problems along a complex contour around the eigenfrequency of interest instead of solving the nonlinear eigenproblem directly. The resulting Riesz projection is normalized within a post-process.

We increase the numerical resolution by increasing the finite element degree up to p=5. Figures 8(c)-(d) show the relative error (see above) of $\tilde{\omega}$ and $\tilde{V}$ for the nanocube and nanocylinder. The relative errors are plotted over the number of unknowns for the scattering solver. We choose three integration points for the contour integration, i.e., three scattering problems are solved for a given finite element degree $p$. The quasi-exact solution is the numerical result obtained with $p = 5$. The unstructured mesh for the nanocube is shown in Fig. 7(a), it contains 40522 tetrahedrons with a minimal edge length of about 0.06 nm and a maximal edge length of about 36 nm. The fine meshing around the rounded corners and edges of the nanocube ensures that the geometry model error is sufficiently small. We have checked that further refinement of the corners changes the numerical results at $p = 4$ (most accurate results in the plots in Fig. 8) only by amounts smaller than the obtained accuracy. E.g., the impact of further refining the mesh on Re $(\tilde{\omega})$ is below $10^{-5}$. At a numerical resolution with $p = 4$ we obtain a resonance wavelength of $\tilde{\lambda} = 665.8529 + 15.57386i$ nm $\pm 0.0184 + 0.00092i$ nm. At a numerical resolution with $p = 5$ we obtain a resonance wavelength of $\tilde{\lambda} = 665.8510 + 15.57396i$ nm.

For the nanocylinder, we consider a two-dimensional cross section of a rotationally symmetric geometry. The unstructured mesh consists of 1086 triangles. At a numerical resolution with $p = 4$ we obtain a resonance wavelength of $\tilde{\lambda} = 663.5622 + 14.68720i$ nm $\pm 0.0207 + 0.00197i$ nm. At a numerical resolution with $p = 5$ we obtain a resonance wavelength of $\tilde{\lambda} = 663.5601 + 14.68740i$ nm.

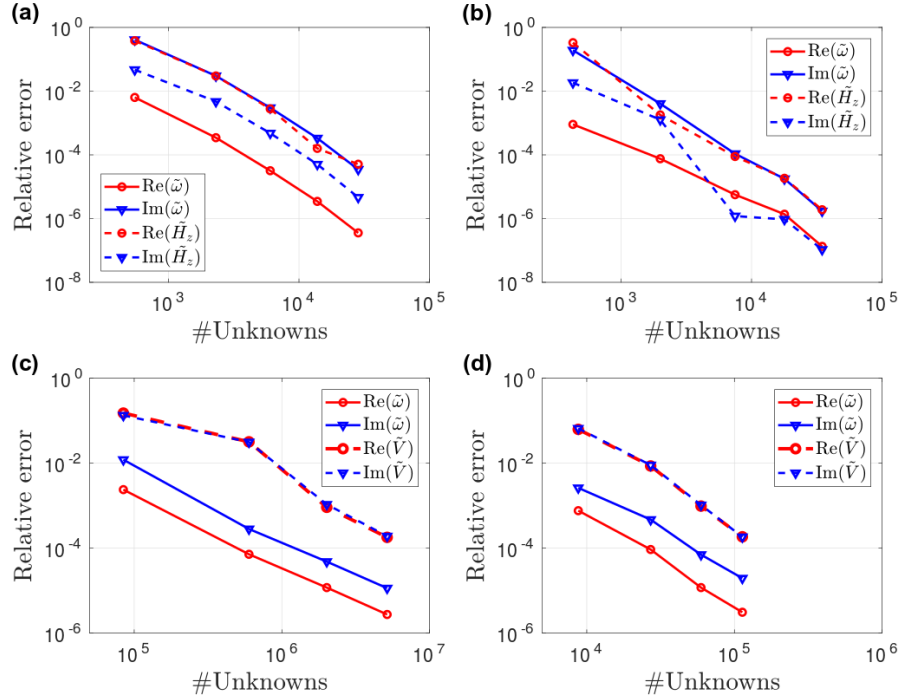

**Fig. 8.** Convergence of FEM1 results: (**a-d**) Relative error of the real and imaginary parts of the eigenfrequency $\tilde{\omega}$, and of the field values, $\tilde{H}_z$ and $\tilde{V}$, as a function of the number of unknowns for benchmark case 1 (**a**), case 2 (**b**), case 3 / nanocube (**c**), case 3 / nanocylinder (**d**).

## 4.4 LCF, Palaiseau

Two different in-house software implemented with MATLAB have been used to solve the three benchmarks. The first one uses the FMM [Moh95] and its aperiodic extension [Sil01,Hug05], and the second one the FEM [Jin14].

### 4.4.1. Fourier Modal Method FMM3 - LCF

The FMM3 used for benchmarks 1 and 2 is a FMM used in transverse magnetic polarization [Moh95,Lal96]. Our implementation includes the S-matrix algorithm [Li96], an accurate computation of the field with Fourier-series expansions [Lal98], and adaptative spatial resolution around air/metal interfaces [Gra99,Val02]. Since the nanocube resonator of benchmark 3 is not a periodic structure, we use for the computation the aperiodic FMM (a-FMM) [Sil01] implemented with complex coordinate transforms [Hug05]. For the cylindrical version of benchmark 3, we use a particular implementation of the a-FMM for body-of-revolution objects [Big14].

In all cases, we compute the QNM eigenfrequencies with the pole-search approach. For the convergence calculation, the guess value for the iterative pole with $N$ Fourier harmonics is the prior result of the calculation with $N - 1$ Fourier harmonics. The initial guess value (first calculation with the smallest value of $N$) is chosen as the real part of the complex eigenfrequency, rounded to the nearest number with two significant digits, e.g. $\omega a / (2\pi c) = 0.23$ for benchmark 1. After each iteration, the QNM field is computed as the field scattered by a dipole source exciting the resonator at the complex frequency of the pole [Bai13].

The following describes some technical issues that are specific to each benchmark. We finally show full convergence curves for the calculations of benchmarks 2 and 3. We also describe the method used to determine the most accurate results and the corresponding relative error.

For the plasmonic crystal, we have slightly reformulated the FMM in order to apply periodic boundary conditions in the $y$ direction instead of the usual outgoing waves conditions, the analytical integration with the S-matrix method being performed along the $y$-axis with a fixed wavevector along $x$, $k_x = 0.5\pi/a$. We calculate the QNM field by inserting a magnetic dipole source in the center of the metallic square. Then we deduce from the symmetries of the system the field of the counter-propagating QNM with $-k_x$. Finally, the field is normalized by calculating the surface integral in Eq. (7) over the surface of one unit cell.

For the grating example, the analytical integration of the FMM is performed along the $y$-axis with a fixed wavevector along $x$, $k_x = 0.2\pi/a$ [Cao02,Lec07]. The rest of the computation is the same as for Benchmark 1. In principle, the use of PMLs in the $y$ direction is not necessary since outgoing boundary conditions are automatically satisfied with the FMM. However, we are using PMLs for normalization purposes, according to Eq. (7).

For the third example, the QNMs are computed by inserting a magnetic dipole source (linearly polarized along the $x$-direction) below the metallic patch in the center of the dielectric layer. After computing the pole, we obtain the QNM field by computing the field scattered by the dipole source at a complex wavelength slightly shifted from the pole value. Unlike benchmarks 1 and 2, the QNMs are then normalized with the *QNMPole* method [Bai13].

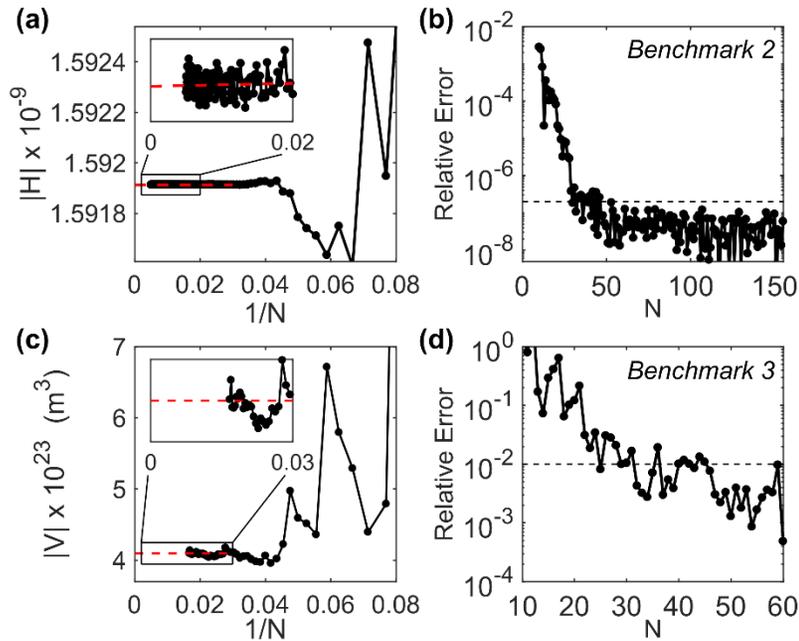

**Fig. 9.** Estimation of the relative error for Benchmarks 2 (**a**)-(**b**) and 3 (**c**)-(**d**). (**a**) Convergence of the normalized field modulus as a function of $1/N$, $2N+1$ being the total number of Fourier harmonics used in the FMM. (**b**) Relative error on the field modulus as a function of $N$. The horizontal dashed line marks an error of $2\times10^{-7}$. (**c**) Convergence of the modulus of the mode volume for Benchmark 3 as a function of $1/N$, $(2N+1)^2$ being the total number of Fourier harmonics used in the aFMM. (**d**) Relative error on the volume modulus as a function of $N$. The horizontal dashed line marks an error of $10^{-2}$. In (**a**) and (**c**), the red dashed lines represent the extrapolated values for $1/N \rightarrow 0$ used to compute the relative errors in (**b**) and (**d**). The insets display a zoom of the convergence curve inside the rectangular box. The vertical scales in the insets in (a) and (b) span over 700 and $1.6\times10^{-24}$, respectively.

To estimate the relative error of our numerical results, we follow a strict procedure, illustrated in Fig. 9 for two examples: the normalized QNM field for Benchmark 2, Figs. 9(a)-(b), and the mode volume for Benchmark 3 (nanocube with 90° corners), see Figs. 9(c)-(d). Figure 9 shows the convergence of a calculated physical quantity ($\widetilde{\omega}$, $\widetilde{V}$...) as a function of the number of Fourier harmonics $N$ used with the FMM or the aFMM. Then, we extrapolate the value of this physical quantity for $1/N \rightarrow 0$. The extrapolation is done by finding a polynomial of degree 0 that best fits the convergence curve in a least-squares sense, see the red dashed lines in Figs. 9(a) and (c). The fit is performed with the numerical data comprised between the smallest $1/N$ value and a larger value ($1/N = 0.02$ in (a) and 0.03 in (c)) for which convergence is assumed to be reached. As shown in the insets of Figs. 11(a) and (c), convergence is reached with a numerical noise. The dashed lines in Fig. 9(b) and (d) highlights that, for $N \geq 50$, a relative error as small as 2x10⁻⁷ is reached for the 2D structure of Benchmark 2 and a relative error of 10⁻² is reached for the 3D structure of Benchmark 3.

### 4.4.2. Finite Element Method FEM2 - LCF

FEM2 is based on edge Whitney elements of order 2 with triangular and tretrahedral adaptative meshes. The QNMs of all three benchmarks have been calculated with a combination of the auxiliary-field and polynomial formulations presented in Section 2.4.

To unveil the implementation, let us consider a single-pole Lorentz model. First, by introducing the current density $\mathbf{J} = i\omega\varepsilon_0\varepsilon_\infty \frac{\omega_p^2}{\omega^2 - \omega_0^2 + i\omega\gamma}\mathbf{E}$ as an auxiliary field, the propagation equation $\nabla \times \mu^{-1}\nabla \times \mathbf{E} = \omega^2\varepsilon(\omega)\mathbf{E}$ can be recast into a quadratic eigenproblem,

$$\omega^2\begin{bmatrix} \varepsilon_0\varepsilon_\infty & 0 \\ 0 & 1 \end{bmatrix}\begin{bmatrix} \mathbf{E} \\ \mathbf{J} \end{bmatrix} + \omega\begin{bmatrix} 0 & i \\ -i\varepsilon_0\varepsilon_\infty\omega_p^2 & i\gamma \end{bmatrix}\begin{bmatrix} \mathbf{E} \\ \mathbf{J} \end{bmatrix} + \begin{bmatrix} -\nabla \times \mu^{-1}\nabla \times & 0 \\ 0 & -\omega_0^2 \end{bmatrix}\begin{bmatrix} \mathbf{E} \\ \mathbf{J} \end{bmatrix} = 0. \tag{11}$$

The latter is solved by introducing two extra fields, $\mathbf{E}_1 = \omega\mathbf{E}$ and $\mathbf{J}_1 = \omega\mathbf{J}$, see Section 2.4. Note that the fields $\mathbf{E}$ and $\mathbf{E}_1$ are defined over the whole computational domain, whereas the field $\mathbf{J}$ and $\mathbf{J}_1$ are only defined over the subdomains that contains the Lorentz material. The coupled system of Eq. (11) is then converted into its corresponding continuous Galerkin weak-form formulation with second order edge elements. Note that this formulation can also be used for 2D geometries in TM polarization (benchmarks 1 and 2), with $\mathbf{E} = (E_x, E_y, 0)$. The associated matrix equation takes the following form

$$\begin{bmatrix} 0 & 0 & \mathbf{I} & 0 \\ 0 & 0 & 0 & \mathbf{I} \\ \mathbf{K} & 0 & 0 & -[\mathbf{M}(i)]_E^J \\ 0 & [\mathbf{M}(\omega_0^2)]_J^J & [\mathbf{M}(i\varepsilon_0\varepsilon_\infty\omega_p^2)]_J^E & -[\mathbf{M}(i\gamma)]_J^J \end{bmatrix}\begin{bmatrix} \mathbf{E} \\ \mathbf{J} \\ \mathbf{E}_1 \\ \mathbf{J}_1 \end{bmatrix} = \omega\begin{bmatrix} \mathbf{I} & 0 & 0 & 0 \\ 0 & \mathbf{I} & 0 & 0 \\ 0 & 0 & [\mathbf{M}(\varepsilon_0\varepsilon_\infty)]_E^E & 0 \\ 0 & 0 & 0 & [\mathbf{M}(1)]_J^J \end{bmatrix}\begin{bmatrix} \mathbf{E} \\ \mathbf{J} \\ \mathbf{E}_1 \\ \mathbf{J}_1 \end{bmatrix}, \tag{12}$$

where the elements of the matrices $\mathbf{M}(\alpha)$ and $\mathbf{K}$ are given by $M_{ij}(\alpha) = \int_\Omega w_i(\mathbf{r})\,\alpha\,w_j(\mathbf{r})\mathrm{d}^3\mathbf{r}$ and $K_{ij} = \int_\Omega \nabla \times w_i(\mathbf{r})\,\mu^{-1}\nabla \times w_j(\mathbf{r})\mathrm{d}^3\mathbf{r}$. The test functions $w_i(\mathbf{r})$ are the edge basis functions. The notation $[\mathbf{M}(\alpha)]_B^A$ specifies the size of the matrix and the subdomain over which it is defined. $[\mathbf{M}(\alpha)]_E^E$ is a square matrix defined over the whole computational domain while $[\mathbf{M}(\alpha)]_J^J$ is defined over the Lorentz-material subdomains. The matrices

$[\mathbf{M}(\alpha)]_E^I$ and $[\mathbf{M}(\alpha)]_J^E$ are rectangular matrices that relates one field defined over the whole computational domain to another field defined over the subdomains. Note that the matrix in the right-hand side of Eq. (12) is block-diagonal. It is thus well conditioned and the CPU time for its diagonalization can be reduced compared to other possible configurations of the final linear algebraic system.

### 4.5 Nankai Univ.

The FMM1 is implemented with an in-house Matlab software [Liu10] and run on a workstation computer with 32GB RAM and two Intel(R) Xeon(R) X5450 CPUs of 3.00GHz and 2.99GHz. It incorporates correct Fourier factorization rules to improve the convergence [Lal96,Gra96,Li96a,Li97,Li03], real-to-Fourier space transform to avoid the Gibbs oscillation of discontinuous field components [Lal98], adaptive spatial resolution [Gra99] to improve the convergence, mirror symmetries to save computational memory [Bai05], and the scattering-matrix algorithm [Pop14] to ensure the numerical stability. For the nanocylinder example, the technique of matched coordinates [Wei09] is adopted and further extended to aperiodic problems to model the curved boundaries of the nanocylinder along with PMLs.

For benchmarks 1 and 2, the QNMs are solved as waveguide Bloch modes by treating the structures as periodic waveguides along the $x$-direction [Lec07]. The computation consists in computing the propagation constant $k_{x,m}(\omega)$ for a fixed frequency $\omega$ of the $m$th waveguide Bloch mode, and then to iteratively solve a transcendental equation, $f(\omega) = k_{x,m}(\omega) - k_x = 0$, whose solution is the eigenfrequency $\omega = \widetilde{\omega}_m$ of the QNM. Note that this equation is essentially related to the eigenvalue formulation of Eq. (9) or (10). But differently, for the latter $\widetilde{\omega}_m$ is solved directly for a given $k_x$, while for the former $k_{x,m}(\omega)$ is returned with the FBMM solver for a given $\omega$. To solve the transcendental equation, we adopt a simple algorithm based on an iteration process with a linear interpolation of $f(\omega)$ [Ort00,Li14], which converges in less than 4 (resp. 6) iterations for benchmark 1 (resp. 2). The two initial guessed values to start the iteration are $\frac{\omega a}{2\pi c} = 0.23$ and $0.23-10^{-6}i$ for benchmark 1, and $\lambda = 0.65$ µm and $0.65+10^{-6}i$ µm for benchmark 2.

For the benchmark 3, the QNMs are computed with the pole-search approach (see Section 2.4). The driving source is set to be a $z$-polarized electric planar source at the central $x$-$y$ plane of the gap (composed of 4 symmetrically-shifted Gaussian sources with a ≈12nm standard deviation). The poled function is $f(\lambda) = H_x^{-1}(\lambda)$ at the gap center. The two initial guessed values for the pole are $\lambda = 0.65$ µm and $0.65+10^{-6}i$ µm and convergence is obtained in typically six iterations.

The convergence performance of FMM1 is shown in Fig. 10. Like in Section 4.3, the relative error $\delta$ is defined as $\delta(N) = |A(N) - A_{\mathrm{QE}}|/|A_{\mathrm{QE}}|$, where $A_{\mathrm{QE}}$ is the numerical value obtained for the highest truncated rank $N_{max}$. The relative error $\delta_{\mathrm{QE}}$ of $A_{\mathrm{QE}}$ can be approximately estimated as $\delta(N_{max} - 5)$. $\delta_{\mathrm{QE}}$ is seen to be close to or below 1%. For quantities with much smaller imaginary parts than their real parts, such as $\widetilde{\omega}$ and $\widetilde{V}$, $\delta_{\mathrm{QE}}$ of the imaginary parts is always larger than that of the real parts.

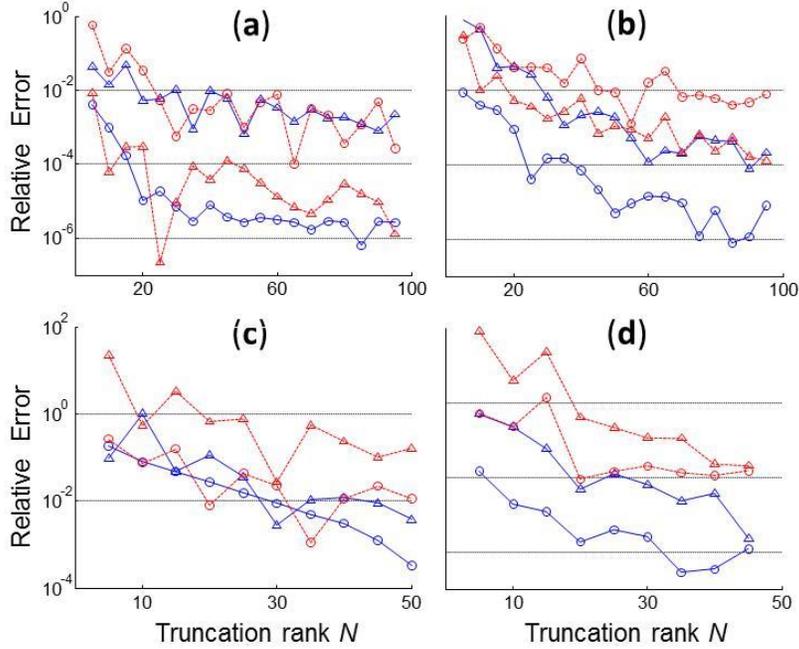

**Fig. 10.** Convergence performance of FMM1 for the plasmonic crystal (**a**), the grating (**b**), the nanocube (**c**) and nanocylinder (**d**). Blue circles: $\text{Re}(\tilde{\omega})$ ; blue triangles : $\text{Im}(\tilde{\omega})$; red circles: $\text{Re}(\tilde{H}_z)$ in (**a-b**) and $\text{Re}(\tilde{V})$ in (**c-d**); red triangles: $\text{Im}(\tilde{H}_z)$ in (**a-b**) and $\text{Im}(\tilde{V})$ in (**c-d**).

### 4.6 Stuttgart Univ.

FMM2 has been used for benchmarks 1 and 2. FMM2 is an in-house implementation of the FMM [Moh95,Lal96], with the final result of the numerical calculations being the scattering matrix that relates incoming with outgoing fields [Whi99,Tik02,Li03]. FMM2 includes the so-called factorization rules [Lal96,Li96a,Li03] and adaptive spatial resolution [Gra99]. Details can be found in [Wei09,Wei11].

The optical resonances are derived from the scattering matrix in an iterative manner. While it is possible to retrieve them as roots of the inverse scattering matrix (i.e., solutions in the absence of any incident field) [Wei11], a general pole ansatz turns out to be more stable [Byk13]. In both cases, a matrix eigenvalue problem is constructed, with the smallest eigenvalues providing guess values for the eigenfrequencies for further iterations. If an eigenvalue is close to zero, the corresponding frequency can be considered as the eigenfrequency of an optical resonance. It has to be emphasized that solving the matrix eigenvalue equation turned out to be much more efficient than using an equivalent root-search algorithm for the corresponding determinant, since all resonances are reduced in the determinant approach to the same single number, so that good guess values are required in order to converge to a certain optical resonance in few iterations. More details can be found in the supplemental material of [Wei16].

A great advantage of the scattering matrix approach for Benchmark 2 is that it does not require PMLs for periodic systems. Furthermore, the outgoing fields in the scattering matrix are given in terms of s- and p-polarized plane waves in the homogeneous top and bottom half spaces, which allows for using analytical normalization schemes [Wei16,Wei17,Wei18]. Here, we have implemented the formulation described in [Wei18].

The following specifications are common to the results in benchmark 1 and 2: The numerical code is implemented in MATLAB, with the Fourier expansion being carried out by fast Fourier transform using a spatial grid with 1024 points.

These points are not equally distributed over one unit cell, but exhibit an increased resolution at the metal-dielectric interfaces. The exact form of the coordinate transformation is taken from [Val02], with the crucial parameter for the resolution increase being G=0. The filling fraction of the metal in the transformed coordinates is 50% of the unit cell. Although the considered benchmarks are effectively one-dimensional periodic with s and p polarization being decoupled, the numerical code calculates solutions for both polarizations at the same time, i.e., it is not optimized for these examples. The normalization of the resonant field distributions is carried out in Fourier space. The method for constructing the spatial fields is described in [Lal98]. In both benchmarks, the truncation rank $N$ has been increased stepwise from 10 to 50 with step size 1. The calculations have been carried out on a desktop PC with 16 GB RAM and an Intel(R) Core(TM) i7-4790 CPU @ 3.60GHz processor.

The plasmonic crystal has no open boundaries. The optical resonances are computed from a modified nonlinear matrix eigenvalue equation that can be derived from the scattering matrix [Cao02]. The eigensolutions have been obtained by the iterative method for finding the roots of a nonlinear matrix eigenvalue equation described in [Wei11]. The initial guess value for the eigenfrequency was $\omega a/(2\pi c) = 0.2310737$. The iterations are stopped when the magnitude of the ratio of the eigenvalue and the frequency drops below $2\times10^{-6}$. The eigenfrequency calculated for the highest accuracy of 101 plane waves is 0.2310737 - 0.000144 $i$.

For the second benchmark, we use the pole method described in the supplemental material of [Wei16].The guess value for the eigenfrequencies was $\omega a/(2\pi c) = 0.7430725 - 0.012661\,i$. The iteration was stopped when the magnitude of the ratio of the eigenvalue and the frequency had dropped below $10^{-7}$ for one eigenvalue. The eigenfrequency calculated for the highest accuracy of 101 plane waves is $0.7430756 - 0.0126606\,i$.

### 4.7 Fresnel Institute, Marseille

For FEM4 – IF, the geometries and meshes are obtained using the GNU software Gmsh [Geu09] and the finite element discretization is handled using GetDP [Dul98]. This open source freeware, developed at the Université de Liège and Université Catholique de Louvain in Belgium, allows to handle the various required basis functions handily. As detailed in [Dem18], several nonlinear eigensolvers from the SLEPc library [Her05] (developed in Universitat Politècnica de València) have recently been added to GetDP in order to tackle problems involving frequency-dispersive materials. Depending on the choice of formulation of the non-linear eigenproblem, GetDP calls linear, general polynomial, or rational eigenvalue SLEPc solvers. These solvers rely on modern Krylov subspace methods. An open source template GetDP 1D grating model is available [http2].

To deal with the unbounded domains, PMLs are used for the domain truncation and to unveil the resonances by rotating the continuous spectrum in the complex plane. In order to take into account the dispersive materials, a suitable representation of the permittivities as functions of the frequency is provided by rational functions. As for the "normalization" of the QNM, we have introduced an exact dispersive quasinormal mode (DQNM) expansion using the Keldysh theorem on eigenfunctions of operators depending on a complex parameter. This expansion relies on the concept of eigentriplets associating eigenvalues to both a right eigenvector (ket) and a left eigenvectors (bra) that may not be in same function spaces. No natural norm can be *a priori* associated to the functions spaces and, moreover, no general biorthogonality relations can be given for the bra-ket pairs. Indeed, the expansion may even not be unique. In the presents numerical benchmarks, we use the conventional expressions of the other teams for comparison purposes.

For benchmarks 1 and 2, the spectra have been computed using four different formulations: (i) using physical auxiliary fields of Eq. (9) coupled with the vector wave equation for the electric field, (ii) using a polynomial eigenvalue formulation of Eq. (11) of the vector wave equation for the electric fiedl, (iii) using a polynomial eigenvalue formulation of Eq. (11) of the scalar wave equation for the z-component of the magnetic field, and (iv) using a rational eigenvalue formulation of the wave equation for the electric field. In the latter case, the linearization is performed internally by the SLEPc solver and the user can only specify the rational function describing the permittivity. The vector electric field formulations are discretized using edge elements (and their higher order generalizations [Web93], order 2 here). The scalar magnetic field formulations are discretized using Lagrange elements of the second order. All electric field formulations lead to the same eigenvalues, up to the machine precision. The scalar magnetic field formulation lead to slightly different results (~$10^{-6}$ in relative precision with respect to the electric field based ones), due to the different basis functions. The benchmark results shown in the paper are those obtained with the scalar magnetic field unknown. In benchmark 1, the initial target value was set to $\widetilde{\omega}a/2\pi c = 0.23 - 0.00014i$. In benchmark 2, the initial target value was set to $\widetilde{\omega}a/2\pi c = 0.75 - 0.01i$.

For benchmarks 3, the results were computed using the rational eigenvalue formulation of the wave equation for the electric field, providing the two dispersive permittivities as rational functions. The two planar symmetries of the structure were taken into account and one fourth of the structure was meshed. In this benchmark, the initial target value was set to $\widetilde{\lambda} = 670 + 15i$.

# 5. Comparison and summary

Benchmarks have a long tradition in computational electromagnetism. They are issued for different purposes, for comparison of modeling approaches and verification of computer codes [Bie06,Las18b], elucidation of controversial interpretations of experimental results [Lal07], or prediction of important figures of merit of optical devices [Cty02].

Since optical resonances play an important role in nanophotonics and impact many areas of modern photonics, the computation of resonator eigenmodes is becoming a major issue. For non-dispersive materials, QNMs can be computed very efficiently with several numerical approaches [Las18b] including commercial software [COMSOL]. Taking into account dispersive materials is more challenging since the eigenproblem becomes nonlinear. This is the issue we have considered here for three different resonator geometries made of highly-dispersive materials, i.e., metals at optical frequencies. The third geometry, composed of two different metals described by permittivity models with a total number of six poles, is exemplary on this point.

For every geometry, we have benchmarked two important physical quantities, the eigenfrequency $\widetilde{\omega}$ and the mode volume $\widetilde{V}$. The latter is directly related to the QNM normalization, which is a crucial issue for reconstruction problems [Lal18]. We have benchmarked several computational approaches, the classical pole-search approach and other more novel ones relying on auxiliary-field or polynomial formulations, and several numerical methods, finite differences, finite elements and Fourier expansions (see Table 1). All three approaches require a description of the material permittivity with an analytical function. The pole-search approach is very general and can be applied with any function describing the dispersion, while the two other approaches require that the permittivity be cast into the specific form of an algebraic fraction between two polynomials of the frequency.

The results of the three benchmarks are summarized in Tables 2-4,6 and in Figs. 3 and 6, which illustrate the convergence performance and wall-clock time as the numerical resolution is increased. Concerning the accuracy of the computed $\widetilde{\omega}$ and $\widetilde{V}$, the conclusion is clear: the different approaches have virtually identical performance, the numerical deviations and wall time differences in the Tables and Figures solely depend on the numerical methods themselves. Some achievements in terms of accuracy are impressive, some others suffer from known issues related to the suitability of the method to the geometry. It is for instance well known that FD schemes have difficulties to handle geometries with curved surfaces, especially for metals. When examining the different data, one should also compare the results with care, since the numerical methods have been benchmarked with different software and computers, the wall-clock time depends on the initial guess value for pole-search approaches, and on subtle numerical details. For instance, the convergence performance of FEMs or (a)FMMs strongly depends on the efforts taken for optimizing the mesh or the adaptive spatial coordinate transform.

Additionally, when analyzing the agreement between all the results, one should put them into perspective, considering that nanoresonators are extremely sensitive structures. For instance, the $\tilde{\omega}$ and $\tilde{V}$ are extremely sensitive to the tiny gap width and length for the nanocube resonators [Aks14], and even the largest numerical deviations observed for the nanocube are well below experimental uncertainties. Aside from the achieved accuracy, by evidencing that several methods can efficiently predict the $\tilde{\omega}$ and $\tilde{V}$ of electromagnetic resonators, we expect to establish reliable standards for QNM computation and normalization. We expect that this standardization will help further developments of modal approaches for analyzing electromagnetic resonances.

## Acknowledgements

The benchmark is motivated by the project RESONANCE ANR-16-CE24-0013 that gathers four French groups with the objective of developing a QNM theory similar to that of normal mode theory of optical waveguides. PL and WY acknowledge the support from the LabEx LAPHIA. PL and AG acknowledge the support of Délégation Générale de l'Armement (DGA) and of INRIA. TW acknowledges support from DFG SPP1839, the VW Foundation, and the MWK Baden-Württemberg. HL acknowledges the support from the National Natural Science Foundation of China (NSFC) (61775105, 11504270) and 111 Project (B16027). SB and FB acknowledge support from Einstein Foundation Berlin within the framework of MATHEON (ECMath, project OT9) and from the European Union's Horizon 2020 research and innovation program under the project 17FUN01 "BeCOMe" of the EMPIR initiative. JZ and RR acknowledge the support by the Dutch Technology Foundation STW ("Good vibrations" project number 14222), which is part of the Netherlands Organisation for Scientific Research (NWO), and which is partly funded by the Ministry of Economic Affairs.